\documentclass[aps,prd,twocolumn,showpacs,preprintnumbers,nofootinbib,amsmath,amssymb,superscriptaddress]{revtex4}

\usepackage{placeins} 
\usepackage{float}
\usepackage{graphicx} 
\usepackage{dcolumn} 
\usepackage{bm} 
\usepackage{amsmath}
\usepackage[normalem]{ulem}

\newcommand{\beq}{\begin{eqnarray}}
\newcommand{\eeq}{\end{eqnarray}}

\newcommand{\ds}{\ensuremath{\displaystyle}}

\def\spose#1{\hbox to 0pt{#1\hss}}
\def\ltapprox{\mathrel{\spose{\lower 3pt\hbox{$\mathchar"218$}}
 \raise 2.0pt\hbox{$\mathchar"13C$}}}

\begin{document}

\title{Curvature of the pseudocritical line in QCD: Taylor expansion matches analytic continuation}
\author{Claudio Bonati}
\email{claudio.bonati@df.unipi.it}
\affiliation{Universit\`a di Pisa, Largo B.~Pontecorvo 3, I-56127 Pisa, Italy}
\affiliation{INFN Sezione di Pisa, Largo B.~Pontecorvo 3, I-56127 Pisa, Italy}
\author{Massimo D'Elia}
\email{massimo.delia@unipi.it}
\affiliation{Universit\`a di Pisa, Largo B.~Pontecorvo 3, I-56127 Pisa, Italy}
\affiliation{INFN Sezione di Pisa, Largo B.~Pontecorvo 3, I-56127 Pisa, Italy}
\author{Francesco Negro}
\email{fnegro@pi.infn.it}
\affiliation{INFN Sezione di Pisa, Largo B.~Pontecorvo 3, I-56127 Pisa, Italy}
\author{Francesco Sanfilippo}
\email{sanfilippo@roma3.infn.it}
\affiliation{INFN Sezione di Roma3, Via della Vasca Navale 84, I-00146 Roma, Italy}
\author{Kevin Zambello}
\email{kevin.zambello@pr.infn.it}
\affiliation{Universit\`a di Parma and INFN, Gruppo Collegato di Parma, Parco Area delle Scienze 7/A, I-43124 Parma, Italy}

\date{\today} 

\begin{abstract}
We determine the curvature of the pseudo-critical line of 
$N_f = 2+1$ QCD with physical quark masses via Taylor expansion
in the quark chemical potentials. We adopt a discretization based on 
stout improved staggered fermions and
the tree level Symanzik gauge action; the location of 
the pseudocritical temperature is based on 
chiral symmetry restoration.
Simulations are performed on lattices with
different temporal extent ($N_t = 6,8,10$), leading to 
a continuum extrapolated curvature
$\kappa = 0.0145(25)$, which is in very good agreement with
the continuum extrapolation obtained via analytic continuation
and the same discretization, 
$\kappa = 0.0135(20)$. This result eliminates the possible tension
emerging when comparing analytic continuation 
with earlier results obtained via Taylor expansion.
\end{abstract}

\pacs{12.38.Aw, 11.15.Ha,12.38.Gc,12.38.Mh}

\maketitle

\section{Introduction}
\label{intro}

The exploration of the QCD phase diagram is a subject of continuous
experimental and theoretical investigation. One of the main
issues is represented by the determination of the pseudo-critical line
in the $T - \mu_B$ plane separating the low-$T$
phase, characterized by color confinement and chiral symmetry
breaking, from the high-$T$ phase where the so-called 
Quark-Gluon Plasma (QGP)
is thought to be realized.
Lattice QCD simulations at non-zero $\mu_B$ are still hindered 
by the well known sign problem, however various methods
are already effective to circumvent the
problem at least for small $\mu_B$, where 
the pseudo-critical line can be approximated at the lowest 
order in a Taylor expansion in $\mu_B^2$
\begin{equation}\label{corcur}
\frac{T_c(\mu_B)}{T_c}=1-\kappa \left(\frac{\mu_B}{T_c}\right)^2\, +\, 
O(\mu_B^4)\, .
\end{equation}
The curvature
$\kappa$ of the pseudo-critical line
has been determined on the lattice both by analytic continuation \cite{alford,lomb99,fp1,dl1,azcoiti,chen,cea_other,Wu:2006su,NN2011,cea2009,alexandru,cea2012}, exploiting
results obtained at imaginary chemical potentials, and by Taylor 
expansion \cite{tay1,tay2,tay3,tay4},
i.e.~by suitable combinations of expectation values determined 
at zero chemical potential.
The pseudo-critical line for small values 
of $\mu_B$ has been investigated also by continuum approaches to the QCD phase 
diagram (see, e.g., Refs.~\cite{Schaefer:2004en,Braun:2011iz,Fischer:2012vc,Pawlowski:2014zaa,Fischer:2014ata}).

Recently, various lattice investigations have led to 
a determination of $\kappa$ by analytic continuation for QCD 
with physical or almost physical quark masses~\cite{crow,corvo2,ntc,ccp,ccp2}. 
In particular, Refs.~\cite{corvo2} and \cite{ntc} have provided
continuum extrapolated values for $\kappa$ which are respectively
$\kappa = 0.0135(20)$ and $\kappa = 0.0149(21)$. The two studies 
adopted a similar discretization 
(stout improved staggered fermions and the tree level Symanzik gauge action)
and slightly different setups for the quark chemical potentials: 
in Ref.~\cite{ntc}, 
the strangeness neutrality 
condition reproduced in the heavy-ion experimental
environment was enforced explicitly by 
tuning the strange quark chemical potential $\mu_s$ appropriately;
in Ref.~\cite{corvo2}, 
instead, 
$\mu_s$ was set to zero while checking at the same time that its
influence on $\kappa$ is negligible (see also 
Ref.~\cite{hegde}). Results obtained by analytic continuation
but adopting a different lattice discretization (HISQ staggered fermions)
have led to similar results~\cite{ccp,ccp2}.

Such results are typically larger than earlier results obtained via 
Taylor expansion~\cite{Kaczmarek2011,Endrodi2011}, reporting
$\kappa \sim 0.006$. 
In particular, Ref.~\cite{Endrodi2011} reported 
a continuum extrapolated value 
$\kappa = 0.0066(20)$ adopting the same discretization
and the same observables (chiral condensate) as in Refs.~\cite{crow,corvo2},
i.e.~a value which is more than
two standard deviations away from the 
result from analytic continuation. As discussed 
in Ref.~\cite{crow}, only a small part of this discrepancy 
can be accounted for by
the different prescriptions used to determine the 
dependence of $T_c$ on $\mu_B$, so that a tension remains.

The agreement between results obtained by the two methods is 
a necessary requirement in order to state that one
has a full control over all systematics involved in analytic continuation
and in Taylor expansion. Therefore, the importance of clarifying 
any possible tension cannot be overestimated; indeed, efforts 
in this direction are already in progress, for instance
by adopting the HISQ staggered discretization~\cite{hegde}.
 
In this study, we present a new continuum extrapolation for the curvature
obtained via Taylor expansion, considering the same stout staggered
discretization adopted in Refs.~\cite{Endrodi2011} and \cite{corvo2}.
In particular, we consider different prescriptions to determine
$\kappa$ via Taylor expansion and the analysis of the renormalized
condensate: for fixed $N_t = 6$, we show that
they provide consistent results; then, exploiting simulations
on different values of $N_t$ ($N_t = 6,8,10$), we are able 
to provide results extrapolated to the continuum limit 
which is in full agreement with
that obtained by analytic continuation.

The paper is organized as follows. In Section~\ref{setup} we provide 
the details regarding the lattice discretization adopted in this study, 
the various prescriptions to determine $\kappa$ 
that we have explored and the observables involved in such prescription.
In Section~\ref{results} we
illustrate 
our numerical results and finally, in Section~\ref{conclusions}
we discuss our conclusions.

\section{Numerical Methods}
\label{setup}

As in our previous studies, Refs.~\cite{crow,corvo2}, 
we have considered a rooted stout 
staggered discretization of the $N_f=2+1$ QCD partition function:
\begin{eqnarray}\label{partfunc}
\mathcal{Z} &=& \int \!\mathcal{D}U \,e^{-\mathcal{S}_{Y\!M}} \!\!\!\!\prod_{f=u,\,d,\,s} \!\!\! 
\det{\left({M^{f}_{\textnormal{st}}[U,\mu_{f,I}]}\right)^{1/4}}
\hspace{-0.1cm}, \\
\label{tlsyact}
\mathcal{S}_{Y\!M}&=& - \frac{\beta}{3}\sum_{i, \mu \neq \nu} \left( \frac{5}{6}
W^{1\!\times \! 1}_{i;\,\mu\nu} -
\frac{1}{12} W^{1\!\times \! 2}_{i;\,\mu\nu} \right), \\
\label{fermmatrix}
(M^f_{\textnormal{st}})_{i,\,j}&=&am_f \delta_{i,\,j}+\!\!\sum_{\nu=1}^{4}\frac{\eta_{i;\,\nu}}{2}\nonumber
\left[e^{a \mu_{f}\delta_{\nu,4}}U^{(2)}_{i;\,\nu}\delta_{i,j-\hat{\nu}} \right. \nonumber\\
&-&\left. e^{- a \mu_{f}\delta_{\nu,4}}U^{(2)\dagger}_{i-\hat\nu;\,\nu}\delta_{i,j+\hat\nu}  \right] ,
\end{eqnarray}
where $\mathcal{S}_{Y\!M}$ is the tree level Symanzik improved gauge 
action~\cite{weisz,curci}, written
in terms of the original link variables through
traces of $n\times m$ 
rectangular loops, $W^{n\!\times \! m}_{i;\,\mu\nu}$,
while the fermion matrix $(M^f_{\textnormal{st}})_{i,\,j}$ is
built up in terms of the two times stout-smeared~\cite{morning} links
$U^{(2)}_{i;\,\nu}$, with an
isotropic smearing parameter $\rho = 0.15$; 
finally, the rooting procedure is
used to remove the residual fourth degeneracy of staggered fermions
(see Ref.~\cite{rooting} for a discussion of possible 
related systematics.).
Note that the quark chemical potentials are treated as external sources, and are set to zero in the simulations.

The quark mass spectrum has been chosen so as to have 
two degenerate light quarks, $m_u = m_d \equiv m_l$.
Standard thermal boundary conditions in the 
temporal direction have been set for bosonic and fermionic
degrees of freedom.
The temperature of the system, $T = 1/(N_t a)$, 
has been changed, for fixed $N_t$, by changing
the lattice spacing $a$ while staying on a line of constant 
physics~\cite{tcwup1,befjkkrs}, 
corresponding to a pseudo-Goldstone pion mass 
$m_{\pi}\simeq 135\,\mathrm{MeV}$ and a strange-to-light mass ratio
$m_s/m_{l}=28.15$. 

\subsection{Physical observables used to locate $T_c$ and their renormalization}

As in Refs.~\cite{crow,corvo2}, the determination of the pseudo-critical
temperature $T_c$ will be based on chiral symmetry restoration,
which is the leading phenomenon in the presence of 
light quark masses. In particular, we will consider
the light quark condensate
\begin{equation}
\langle\bar\psi\psi\rangle_l=\frac{T}{V}\frac{\partial \log Z}{\partial m_l}=
\langle\bar{u}u\rangle+\langle\bar{d}d\rangle\ ;
\end{equation} 
where
\begin{equation}
\langle\bar{\psi}\psi\rangle_f=\frac{T}{V}\frac{\partial \log Z}{\partial m_f}\ ,
\end{equation} 
and $V$ is the spatial volume. The light quark condensate is 
affected by additive and
multiplicative renormalizations and, as in Refs.~\cite{crow,corvo2}, 
we consider two different
renormalization prescriptions. The first one is
\begin{equation} \label{rencond}
\langle\bar{\psi}\psi\rangle_{r1}(T)\equiv\frac{\left[
\langle \bar{\psi}\psi\rangle_l -\frac{\ds 2m_{l}}{\ds m_s}\langle \bar{s}s\rangle\right](T)}{
\left[\langle \bar{\psi}\psi\rangle_l-\frac{\ds 2m_{l}}{\ds m_s}\langle \bar{s}s\rangle\right](T=0)}\ ,
\end{equation} 
and has been introduced in Ref.~\cite{Cheng:2007jq}: the leading additive
renormalization, which is linear in the quark mass, 
cancels in the difference with
the strange condensate, while the multiplicative
renormalization, being independent of T, drops out by normalizing with respect
to quantities measured
at $T = 0$ and at the same UV cutoff.
The second definition, introduced in Ref.~\cite{Endrodi2011},
exploits $T = 0$ quantities to perform the additive 
renormalization and the value of the bare quark mass to take care
of multiplicative ones:
\begin{equation}\label{eq:ren_pres_wupp}
\langle \bar{\psi}\psi\rangle_{r2}=\frac{m_{l}}{m_{\pi}^4}\left(\langle\bar{\psi}\psi\rangle_{l}
-\langle\bar{\psi}\psi\rangle_{l}(T=0)\right)\, .
\end{equation}

The location of $T_c$ is usually defined, in terms of the renormalized 
light condensate, as the point of maximum slope, i.e.~the point
where $\langle\bar{\psi}\psi\rangle^r$ has an inflection point 
as a function of $T$ and 
the absolute value of
$\partial \langle\bar{\psi}\psi\rangle^r/ \partial T$ reaches a maximum. 
Alternatively, one can look at the peak  of the 
chiral susceptibility, i.e.~the maximum of 
$\chi_{\bar\psi\psi}
\equiv {\partial\langle\bar\psi\psi\rangle^r} / {\partial m_l}$.
Studies exploiting analytic continuation have considered both definitions
and then monitored the behavior of $T_c$ as a function of the 
imaginary baryon chemical potential in order to determine $\kappa$.
In our case, the determination of $\kappa$ will be based on the 
matching of derivatives with respect to $T$ and $\mu_B$ computed 
at $\mu_B = 0$; the main error source will be the statistical 
one, which is larger and larger as one considers observables 
representing higher order derivatives.
For this reason,
we will
limit ourselves to the analysis of the renormalized chiral
condensate, which is the lowest derivative, while
starting from a second order derivative like the chiral susceptibility
would be much more difficult.

\subsection{Possible definitions of $\kappa$ via Taylor expansion}

The most natural extension to finite $\mu_B$ of the prescription 
to locate $T_c$ in terms of 
$\langle\bar{\psi}\psi\rangle^r$ is to still
look for an inflection point, i.e.~a point where
$\partial^2 \langle\bar{\psi}\psi\rangle^r/ \partial T^2 = 0$.
In order to understand how $T_c$ will move, at the lowest 
order in $\mu_B$, following this prescription, we need to consider
a Taylor expansion of 
$\langle\bar{\psi}\psi\rangle^r (T)$ 
\beq
\langle \bar{\psi} \psi \rangle^r(T, \mu_B) 
= A(T) + B(T) \mu_B^2 + O(\mu_B^4)
\eeq
where 
\beq
A(T) &\equiv& 
\langle \bar{\psi} \psi \rangle^r(T, 0)
\nonumber \\ 
B(T) &\equiv& 
\frac{\partial \langle \bar{\psi} \psi \rangle^r}{\partial ( \mu_B^2 )} (T,0) \, .
\eeq
The prescription is then to require
\begin{eqnarray}
\label{prescr2}
    0 & = & \frac{ \partial^2 \langle \bar{\psi} \psi \rangle^r}{\partial T^2}
(T, \mu_B) = A''(T) + B''(T) \mu_B^2 \\
      & = & A''(T_c) + A'''(T_c)t + ( B''(T_c) + B'''(T_c)t ) \mu_B^2 \mbox{ . } \nonumber
\end{eqnarray}
where the quantities $A''$, $B''$,  $A'''$ and $B'''$
represent second and third order 
derivatives of $A(T)$ and $B(T)$ with respect
to $T$, $t \equiv T - T_c$ and we have 
performed a lowest order Taylor expansion 
around the pseudo-critical temperature
at $\mu_B = 0$, $T_c$.
Solving Eq.~(\ref{prescr2}) for $t$ one obtains
\begin{equation}
    t = \frac{- B''(T_c)}{A'''(T_c) + B'''(T_c) \mu_B^2} \mu_B^2 = - \frac{B''(T_c)}{A'''(T_c)} \mu_B^2 + O(\mu_B^4) \mbox{ , }
\end{equation}
and finally, following the definition 
of $\kappa$ in Eq.~(\ref{corcur}), one obtains
\beq
\label{defkappa2}
    \kappa 
&=& \frac{B''(T_c)}{A'''(T_c)}T_c \\ &=& 
\frac{   \frac{\partial^2}{\partial T^2} ( \frac{\partial \langle \bar{\psi} \psi \rangle^r(T,\mu_B)}{\partial (\mu_B^2)}|_{\mu_B=0} ) |_{T=T_c}   }
                                                     {   \frac{\partial^3}{\partial T^3} \langle \bar{\psi} \psi \rangle^r(T,0) |_{T=T_c}   }\, T_c 
\, . \nonumber 
\eeq
From a practical point of view, Eq.~(\ref{defkappa2}) means
that one needs to evaluate both the renormalized condensate and 
its $\mu_B^2$-derivative
as a function of $T$ around $T_c$, and that
must be done 
with enough 
precision so that, after a suitable interpolation,
one is able to compute numerically
their third
and second order derivatives with respect to $T$ at $T_c$.

As we shall see, the program above has to face the low 
statistical accuracy attainable with reasonable statistics,
in particular when evaluating the $\mu_B^2$-derivative
and especially on the lattices with higher values of 
$N_t$, which are necessary to take the continuum extrapolation.
For this reason, alternative prescriptions for $\kappa$ have been
adopted in the literature. For instance, in Ref.~\cite{Endrodi2011}, 
the pseudo-critical temperature at finite $\mu_B$ is defined as the temperature
where the renormalized condensate attains the same value as at $T_c$ 
for $\mu_B  = 0$, i.e.:
\begin{equation}
    \langle \bar{\psi}\psi \rangle^r (T, \mu_B^2)|_{T=T_c(\mu_B^2)}  \equiv \langle \bar{\psi}\psi \rangle^r (T_c, 0) \mbox{ . }
\label{eq:def_k_endrodi}
\end{equation}
Then, by definition, the differential $d\langle \bar{\psi}\psi \rangle$ 
must vanish along the curve $T_c(\mu_B)$ 
\begin{equation}
       d\langle \bar{\psi}\psi \rangle^r = \frac{\partial \langle \bar{\psi}\psi \rangle^r}{\partial T}\Big\rvert_{\substack{\mu_B=0}} dT + \frac{\partial \langle \bar{\psi}\psi \rangle^r}{\partial (\mu_B^2)}\Big\rvert_{\substack{\mu_B=0}} d(\mu_B^2) = 0
\end{equation}
so that one obtains
\begin{equation}
    \kappa = - T_c  \frac{dT_c}{d(\mu_B^2)} = T_c \frac{ \frac{\partial \langle \bar{\psi}\psi \rangle^r}{\partial (\mu_B^2)}|_{\mu_B = 0, T=T_c} }{ \frac{\partial \langle \bar{\psi}\psi \rangle^r}{\partial T}|_{\mu_B = 0, T =
          T_c} } \mbox{ . }
\label{defkappa1}
\end{equation}
The advantage of the expression in Eq.~(\ref{defkappa1}) with respect
to that in Eq.~(\ref{defkappa2}) is twofold:
one needs to estimate just the first derivative of the renormalized
condensate at $T_c$, which is more precise and stable against
the choice of the interpolating function than the third one, and one does 
not need to know the dependence of 
${\partial \langle \bar{\psi}\psi \rangle^r}/{\partial (\mu_B^2)}$ on $T$,
but just its value at $T_c$.
However, the prescription is debatable, since there is 
no strict reason that the condensate should stay constant in value 
at $T_c$. However, numerical studies at imaginary chemical 
potential~\cite{crow} have shown that it gives results
for $T_c$ which are compatible, within errors, with those obtained 
by looking at the inflection point.

In the following we will consider both definitions, 
and refer to them
as $\kappa_1$, Eq.~(\ref{defkappa1}),  
and $\kappa_2$, Eq.~(\ref{defkappa2}).
As we shall see, a detailed
comparison between the two definitions will be possible only on 
$N_t = 6$ lattices, where they will give compatible results,
while on lattices with larger $N_t$ the statistical errors
attained for $\kappa_2$ will make it practically useless, so that
our present continuum extrapolation will be based on
$\kappa_1$ alone. Yet, $\kappa_1$ is exactly the prescription
adopted in Ref.~\cite{Endrodi2011}, so that a strict comparison
will be possible with the results reported there.

Notice that other prescriptions can be found in the literature, 
which will not be explored in this study.
For instance, the determination 
reported in Ref.~\cite{Kaczmarek2011} 
(see also Refs.~\cite{Ejiri2009,Laermann2013,hegde})
assumes a behavior for the pseudocritical
temperature which is dictated by the critical scaling
around the possible second order point in the 
$O(4)$ universality class located at $m_{l}=0$.

\subsection{Observables needed to determine $\kappa$ and setup of 
chemical potentials}

Apart from the renormalized chiral condensate, which has been already 
defined above, the other quantity needed for our study is its
derivative with respect to $\mu_B^2$. Looking at
Eqs.~(\ref{rencond}) and (\ref{eq:ren_pres_wupp}) one realizes that
such derivative is trivially obtained combining the derivatives of the 
finite temperature flavor condensates with respect to $\mu_B^2$, 
since zero temperature quantities are independent of $\mu_B$ 
around $\mu_B = 0$. Therefore, we need to compute
\begin{eqnarray}
        \frac{\partial \langle \bar{\psi}\psi \rangle_f}{\partial (\mu_B^2)}\Big\rvert_{\substack{\mu_B=0}} &=& \frac{1}{2}
         \frac{\partial^2 \langle \bar{\psi}\psi \rangle_f}{\partial \mu_B^2}\Big\rvert_{\substack{\mu_B=0}} \\
&=&            \frac{1}{2} \langle (n^2 + n') \bar{\psi}\psi_f \rangle - 
\frac{1}{2}
\langle n^2 + n' \rangle \langle
            \bar{\psi}\psi_f \rangle \nonumber \\
        &  & +\ \frac{1}{2} 
\langle 2 n \bar{\psi}\psi_f' + \bar{\psi}\psi_f'' 
\rangle \mbox{ , } \nonumber
\end{eqnarray}
where the relevant operators entering previous expression are defined
as  
\begin{eqnarray}
\label{deftraces}
    \bar{\psi}\psi_f & = & \frac{T}{V} \frac{1}{4} Tr[M_f^{-1}] \, ,
\\
    \bar{\psi}\psi_f' & = & \frac{T}{V} \frac{1}{4} Tr[-M_f^{-1} M_f' M_f^{-1}] \mbox{ , } \nonumber \\
    \bar{\psi}\psi_f'' & = & \frac{T}{V} \frac{1}{4} Tr[2M_f^{-1} M_f' M_f^{-1} M_f' M_f^{-1} 
-\, M_f^{-1} M_f'' M_f^{-1}] \mbox{ , } \nonumber \\
    n & = & \sum_{f=uds} \frac{1}{4} Tr [ M_f^{-1} M_f' ] \mbox{ , } \nonumber \\ 
    n' & = &  \sum_{f=uds} \frac{1}{4} Tr [ M_f^{-1} M_f'' - M_f^{-1} M_f' M_f^{-1} M_f' ] \mbox{} \nonumber
\end{eqnarray}
while $M_f'$ and $M_f''$ represent first and second derivatives
of the fermion matrix, defined in Eq.~(\ref{fermmatrix}), with respect
to $\mu_B$, computed at $\mu_B = 0$.

The way in which such derivatives,
$M_f'$ and $M_f''$, are actually taken depends on
the quark flavor $f$ and specifies our setup of quark chemical potentials.
In particular, as in Refs.~\cite{Endrodi2011,crow,corvo2}, we 
set $\mu_u = \mu_d = \mu_l = \mu_B/3$ and $\mu_s = 0$. Therefore, for 
the strange flavor we 
have $M_s' = M_s'' = 0$, so that
$\bar{\psi}\psi_s'$ and $\bar{\psi}\psi_s''$ trivially vanish.
Instead for $f = u,d$,
considering Eq.~(\ref{fermmatrix}) and taking into account that 
$\partial/\partial \mu_B = 
(1/3) \partial / \partial \mu_l$, we have
\beq
{M_f'}_{\ i,\,j} = \frac{\eta_{i;\,4}}{6} 
\left[U^{(2)}_{i;\,4}\delta_{i,j-\hat{4}} 
 + U^{(2)\dagger}_{i-\hat4;\,4}\delta_{i,j+\hat 4}  \right] ,
\eeq
\beq
{M_f''}_{\ i,\,j} = \frac{\eta_{i;\,4}}{18} 
\left[U^{(2)}_{i;\,4}\delta_{i,j-\hat{4}} 
 - U^{(2)\dagger}_{i-\hat4;\,4}\delta_{i,j+\hat 4}  \right] \, .
\eeq
All traces appearing in Eq.~(\ref{deftraces}) have been computed,
as usual, by multiple noisy estimators, paying attention not to combine
the same random vectors when estimating product of traces in order
to avoid cross-correlations.

\begin{table}[htbp]
\begin{center}
    \begin{tabular}{|c|c|c|c|c|}
    \hline
    $\beta$ & $m_s$ & $a$ $(fm)$ & Lattice\\
    \hline
    $3.49$ & $0.132$ & $0.2556$ & $N_t = 6$\\
    $3.51$ & $0.121$ & $0.2425$ & $N_t=6$\\
    $3.52$ & $0.116$ & $0.2361$ & $N_t=6$\\
    $3.525$ & $0.11350$ & $0.23297$ & $N_t=6$\\
    $3.53$ & $0.111$ & $0.2297$ & $N_t=6$\\
    $3.535$ & $0.10873$ & $0.22663$ & $N_t=6$\\
    $3.54$ & $0.10643$ & $0.2235$ & $N_t=6$\\
    $3.545$ & $0.10419$ & $0.22039$ & $N_t=6$\\
    $3.55$ & $0.10200$ & $0.2173$ & $N_t=6$\\
    $3.555$ & $0.099864$ & $0.21424$ & $N_t=6$\\
    $3.56$ & $0.09779$ & $0.2112$ & $N_t=6$\\
    $3.565$ & $0.095750$ & $0.20820$ & $N_t=6$\\
    $3.57$ & $0.09378$ & $0.2052$ & $N_t=6$\\
    $3.58$ & $0.08998$ & $0.1994$ & $N_t=6$\\
    $3.60$ & $0.08296$ & $0.1881$ & $N_t=6,8$\\
    $3.62$ & $0.07668$ & $0.1773$ & $N_t=8$\\
    $3.63$ & $0.07381$ & $0.1722$ & $N_t=8$\\
    $3.635$ & $0.07240$ & $0.1697$ & $N_t=8$\\
    $3.64$ & $0.07110$ & $0.1672$ & $N_t=8$\\
    $3.645$ & $0.06978$ & $0.1648$ & $N_t=8$\\
    $3.655$ & $0.06731$ & $0.1601$ & $N_t=8$\\
    $3.66$ & $0.06615$ & $0.1579$ & $N_t=8$\\
    $3.665$ & $0.06500$ & $0.1557$ & $N_t=8$\\
    $3.67$ & $0.06390$ & $0.1535$ & $N_t=8$\\
    $3.675$ & $0.06284$ & $0.1514$ & $N_t=8$\\
    $3.68$ & $0.06179$ & $0.1493$ & $N_t=8$\\
    $3.69$ & $0.05982$ & $0.1453$ & $N_t=8$\\
    $3.71$ & $0.05624$ & $0.1379$ & $N_t=8$\\
    $3.74$ & $0.05168$ & $0.1280$ & $N_t=10$\\
    \hline
    \end{tabular}
\end{center}
\caption {List of the bare quark masses and lattice spacings used in our simulations. Bare parameters have been chosen so as to stay on a line of constant
physics with a physical value of the pseudo-Goldstone pion mass, interpolating 
results reported in Refs.~\cite{tcwup1,befjkkrs}. The strange-to-light mass ratio has been fixed to $m_s/m_l = 28.15$. 
}
\label{tab:lcp}
\end{table}

\begin{figure}[t!]
\centering
\includegraphics[scale=0.38]{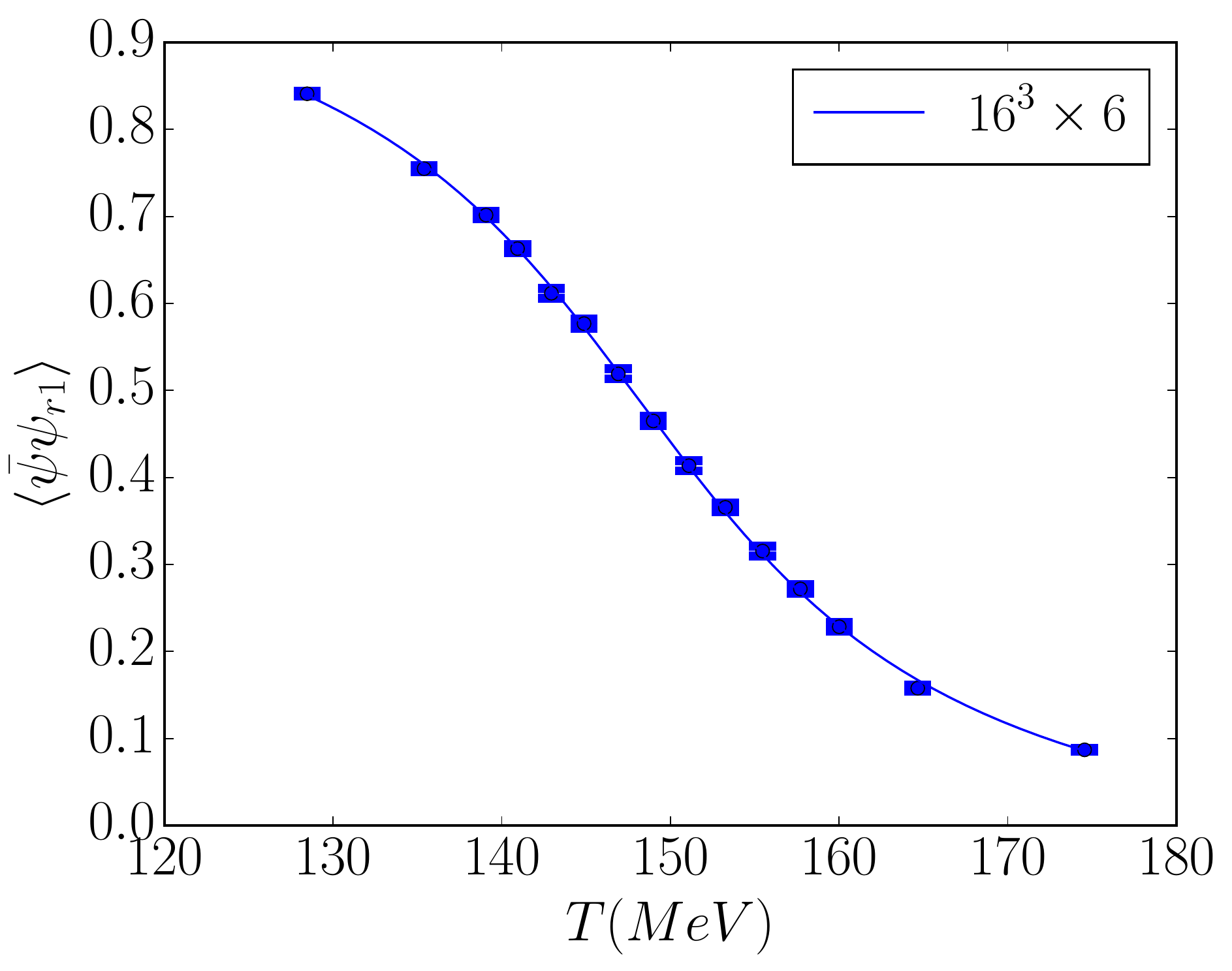}
\includegraphics[scale=0.38]{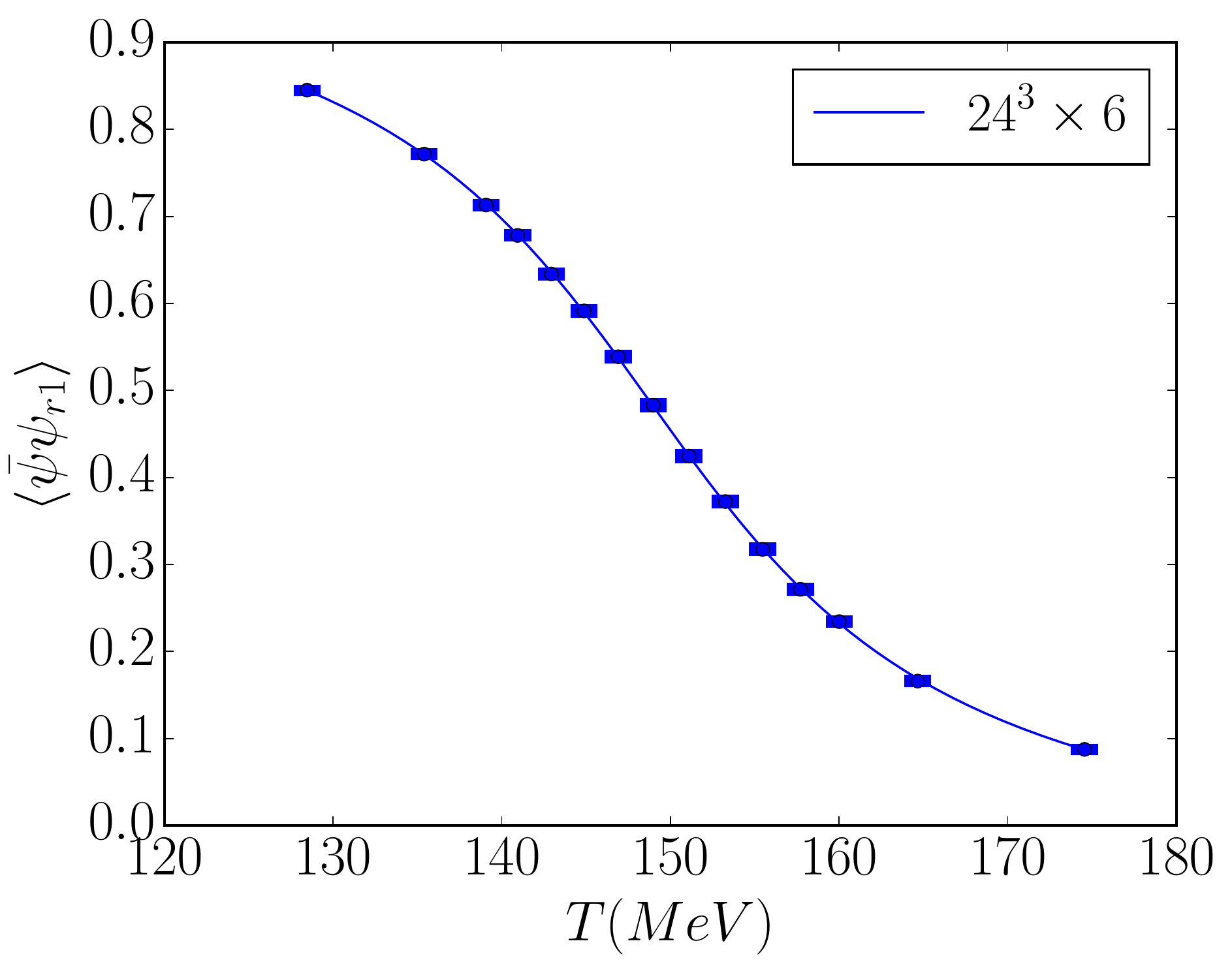}
\includegraphics[scale=0.38]{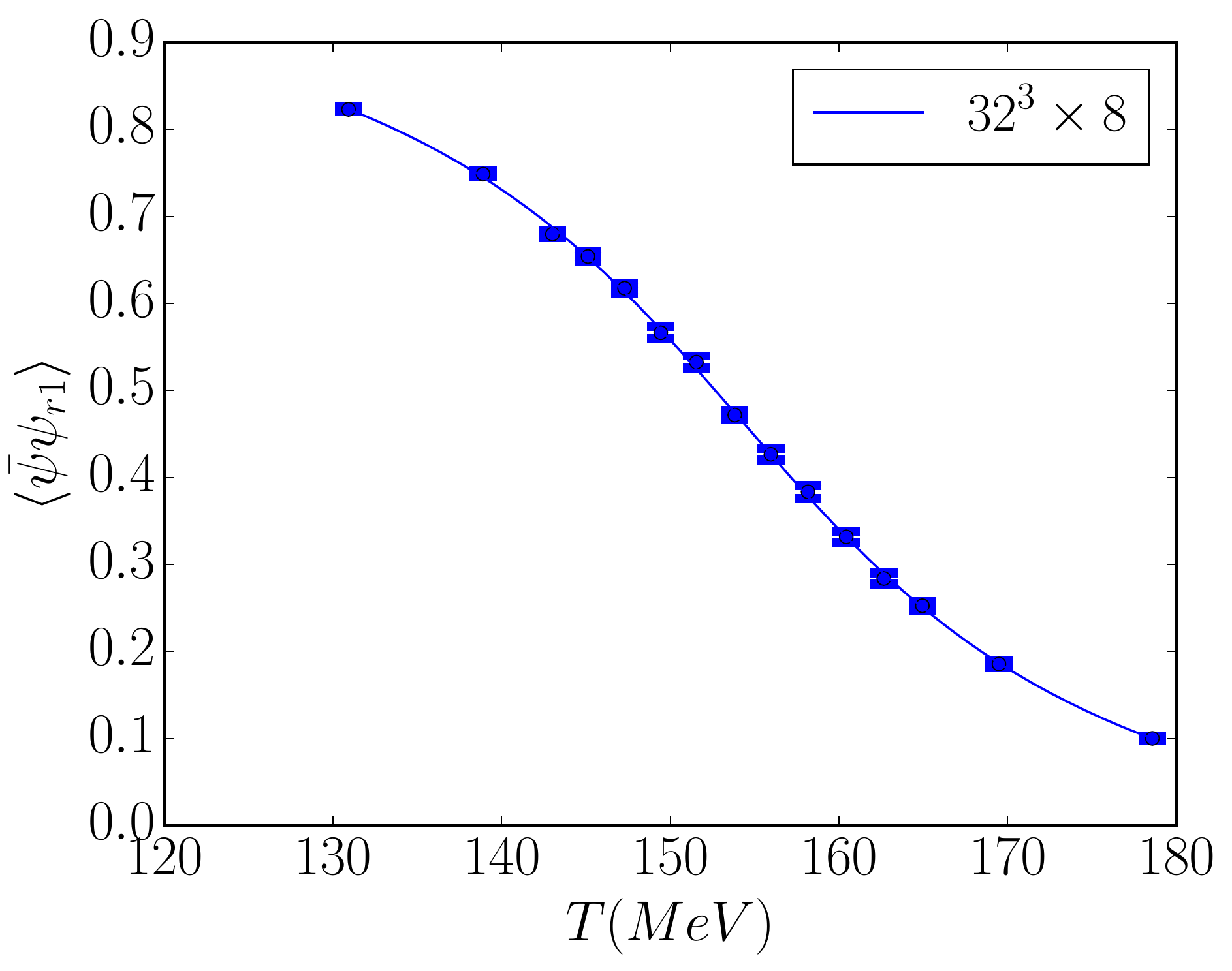}
\caption{Renormalized chiral condensate $\langle \bar{\psi}\psi_{r1} \rangle$
for $N_t = 6$ and $N_t = 8$ lattices.}   
\label{fig:ff_renorm1_all}
\end{figure}

\begin{figure}[t!]
\centering
\includegraphics[scale=0.38]{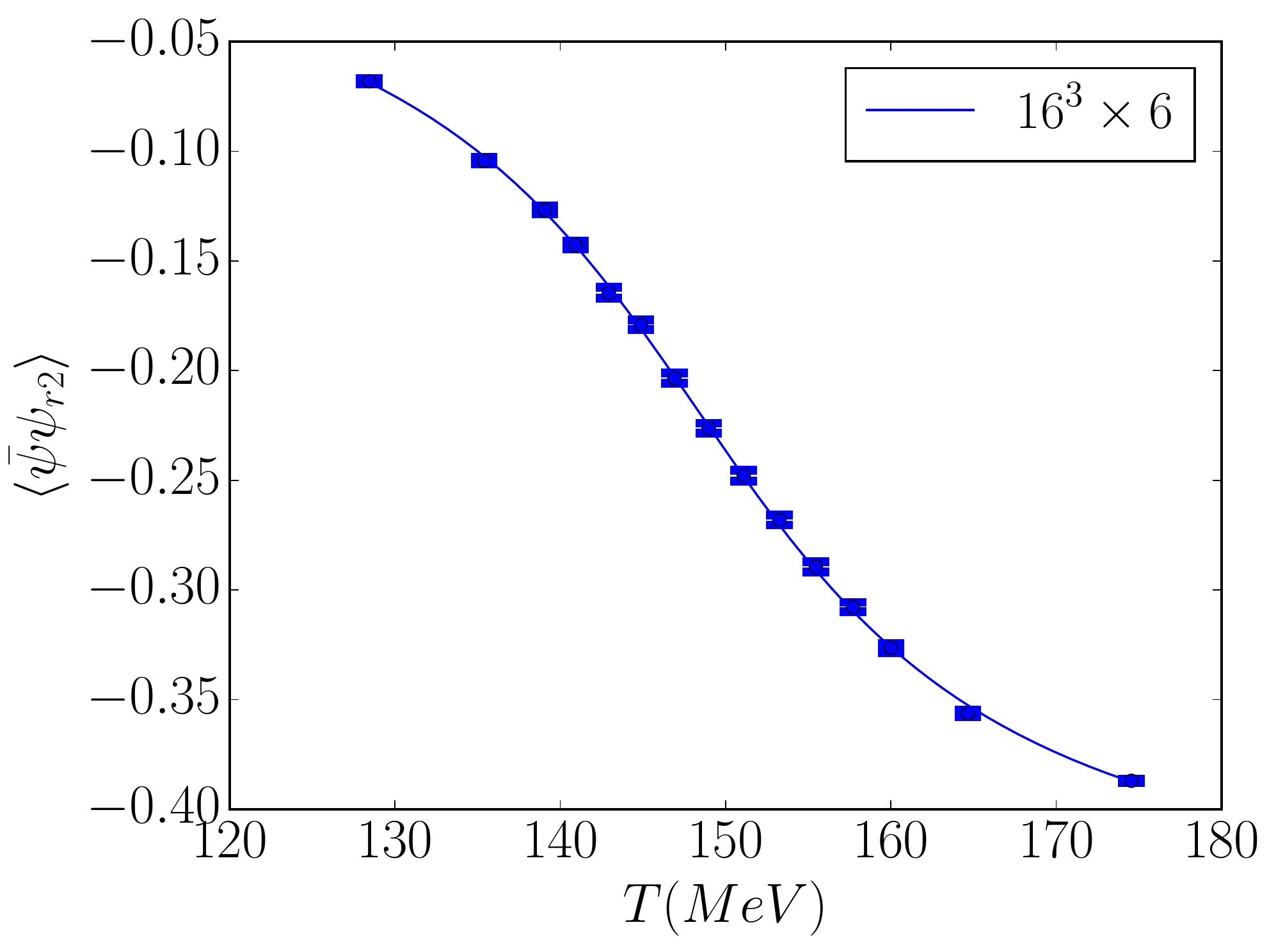}
\includegraphics[scale=0.38]{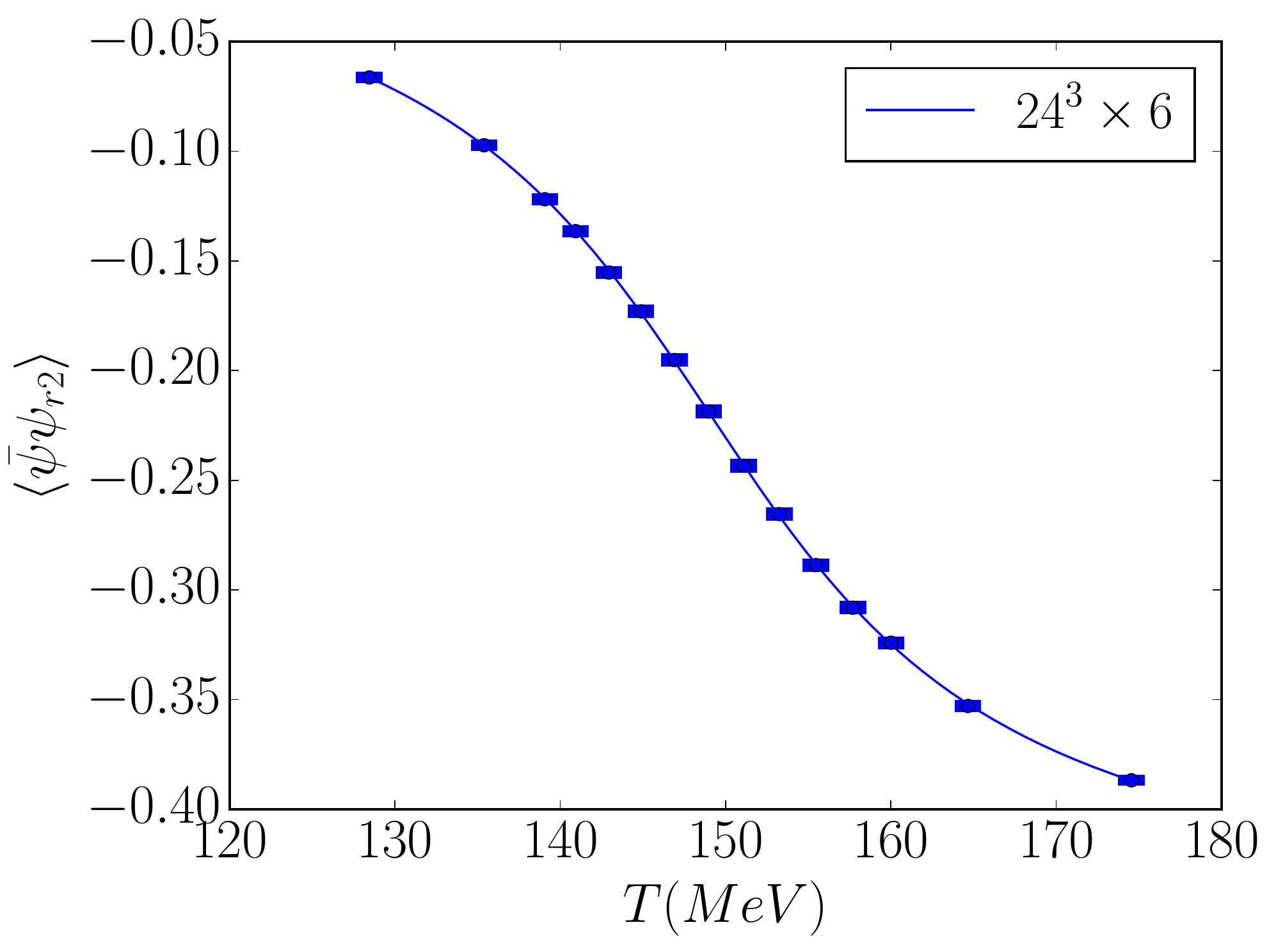}
\includegraphics[scale=0.38]{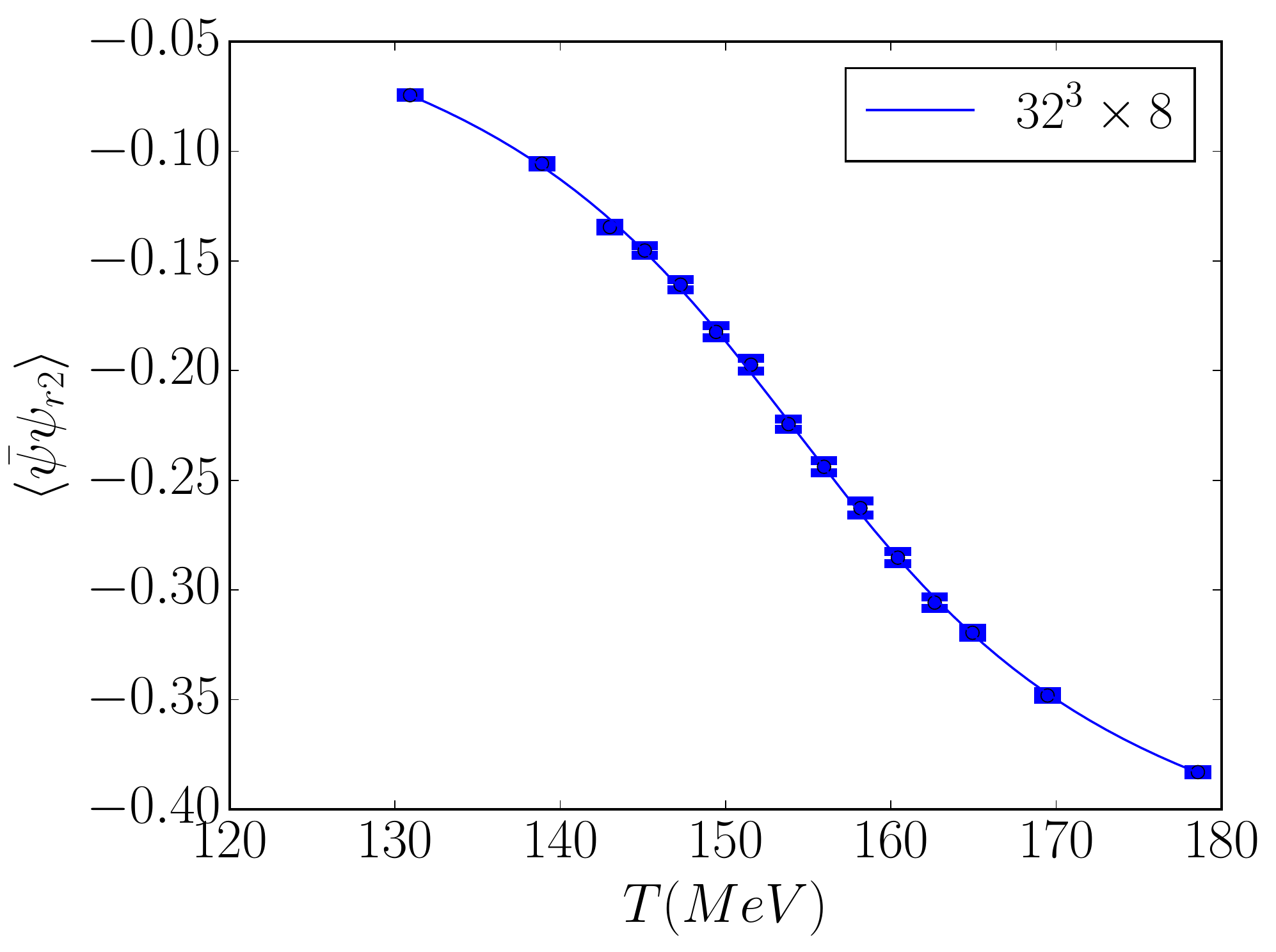}
\caption{Renormalized chiral condensate $\langle \bar{\psi}\psi_{r2} \rangle$
for $N_t = 6$ and $N_t = 8$ lattices.}   
\label{fig:ff_renorm2_all}
\end{figure}

\begin{figure}[t!]
\centering
\includegraphics[scale=0.38]{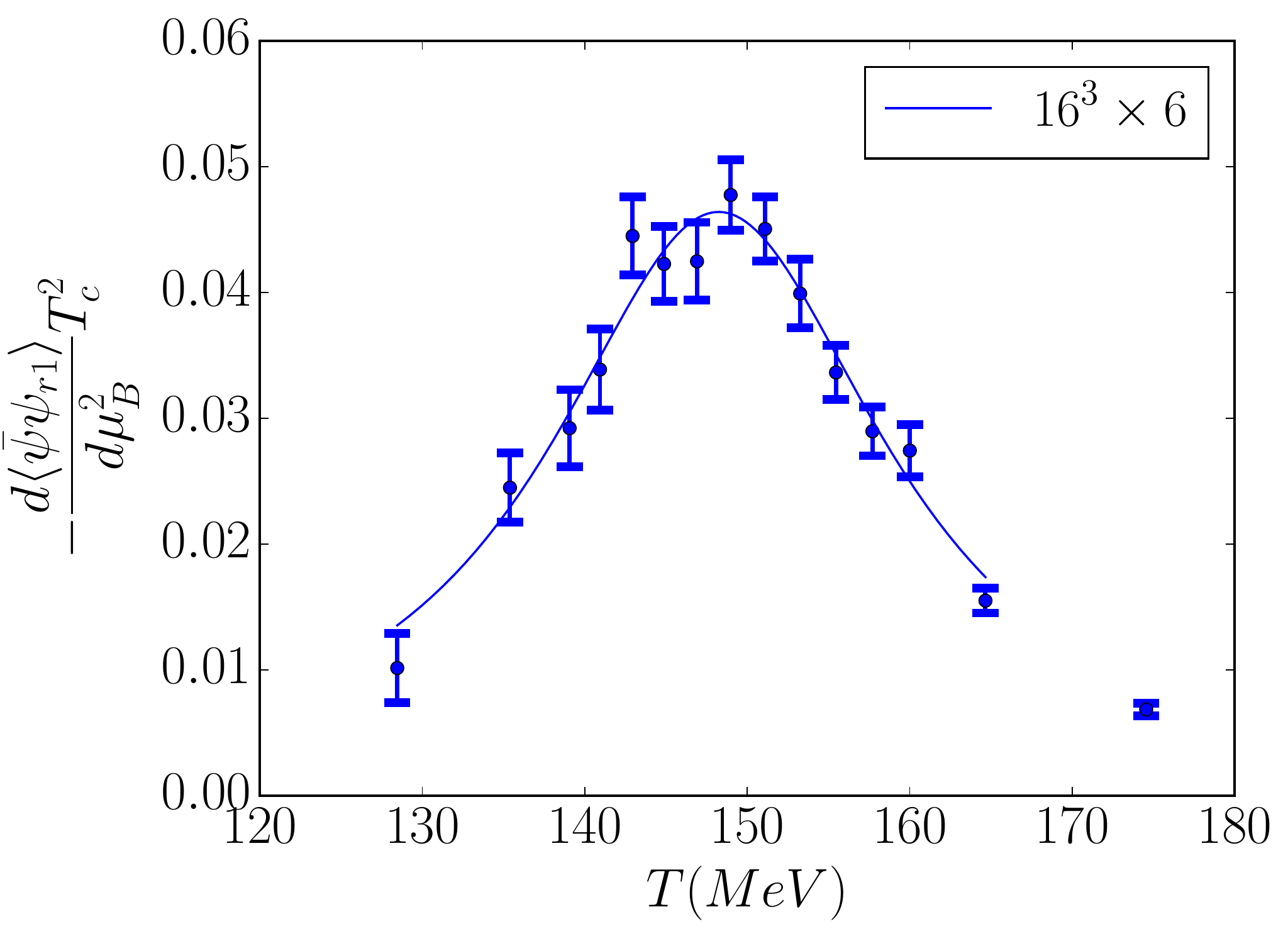}
\includegraphics[scale=0.38]{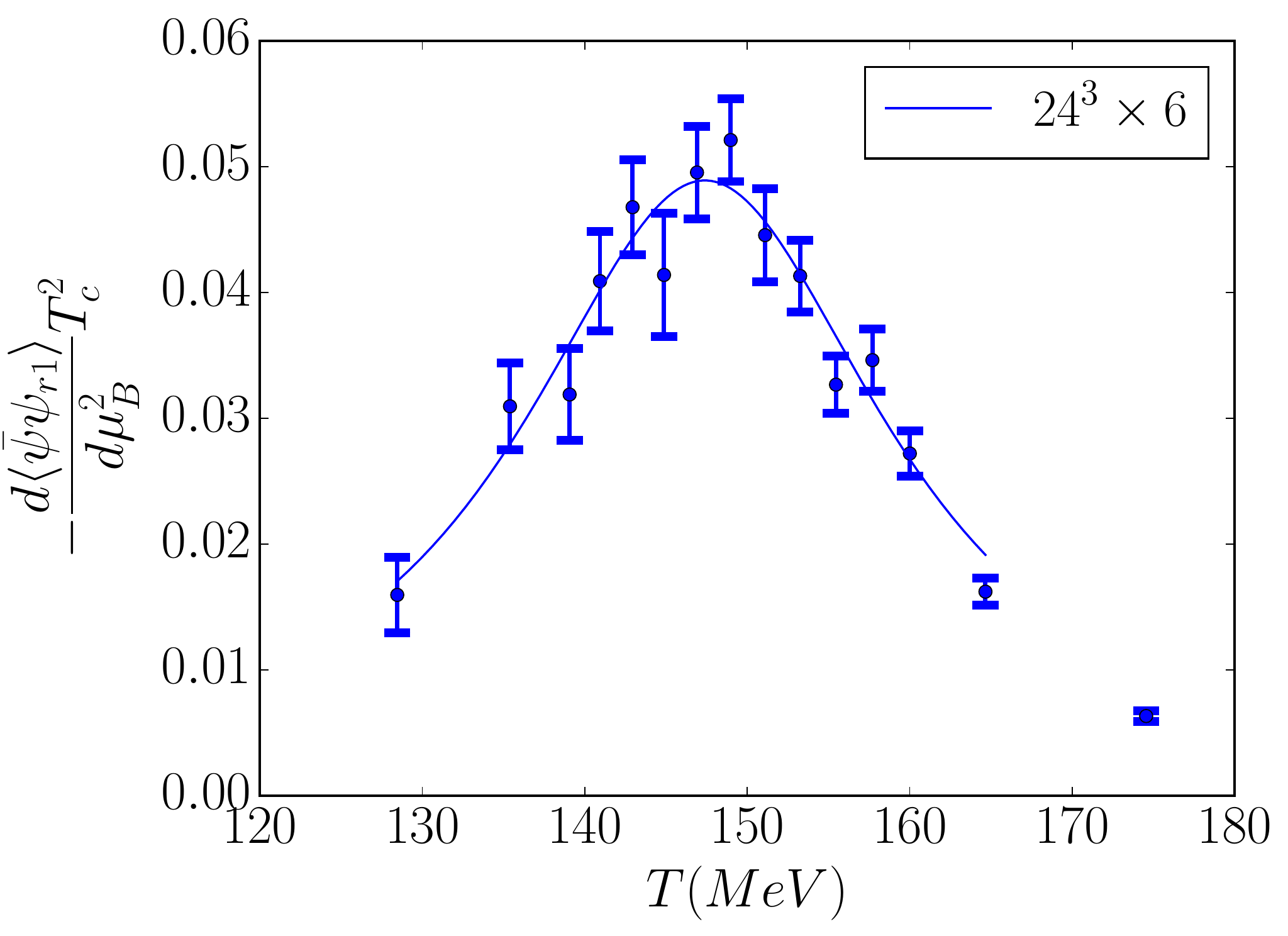}
\includegraphics[scale=0.38]{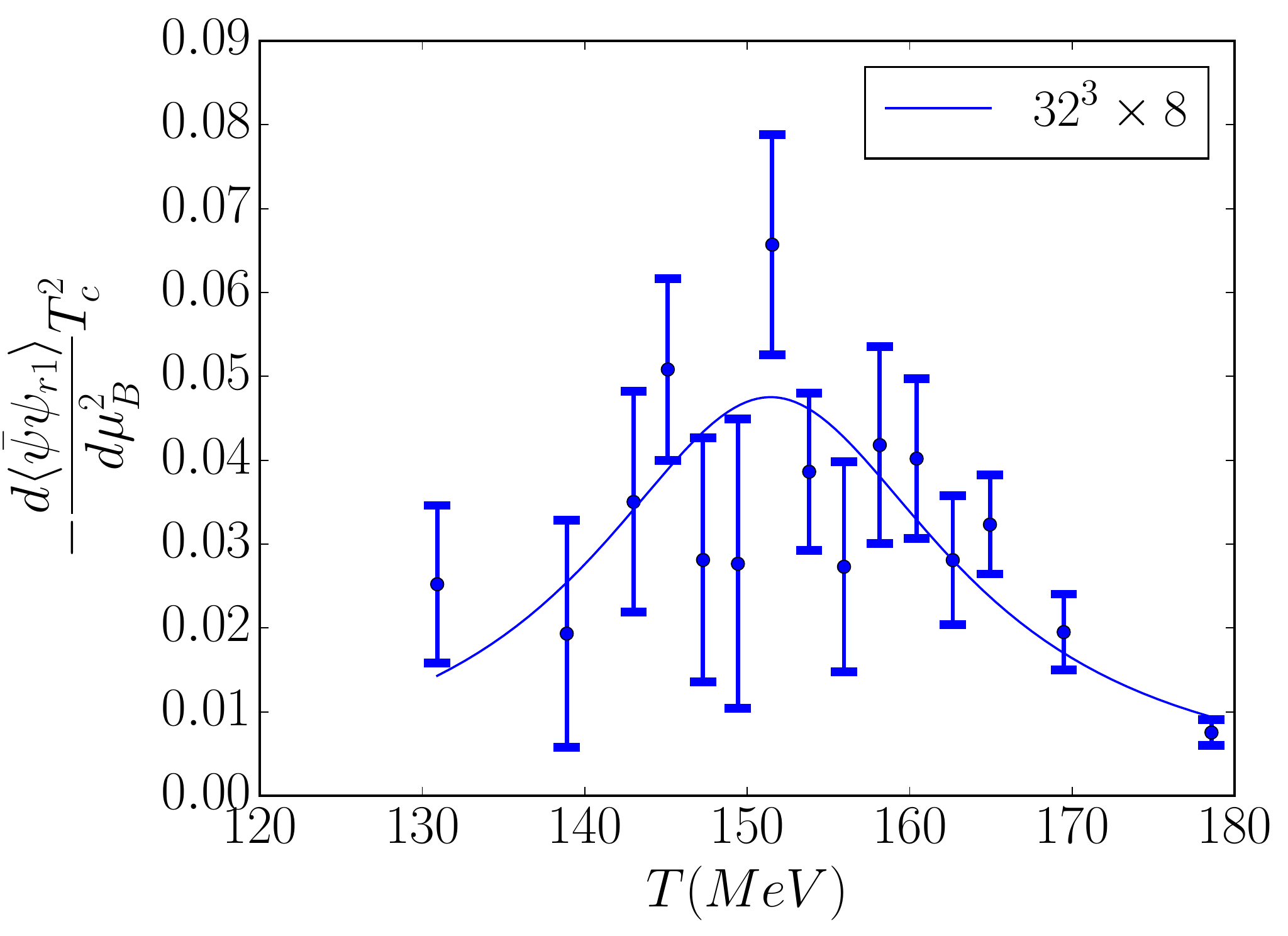}
\caption{First derivative with respect to $\mu_B^2$ of $\langle \bar{\psi}\psi_{r1} \rangle$ for $N_t = 6$ and $N_t = 8$ lattices.}
\label{fig:ffii_renorm1_all}
\end{figure}

\begin{figure}[t!]
\centering
\includegraphics[scale=0.38]{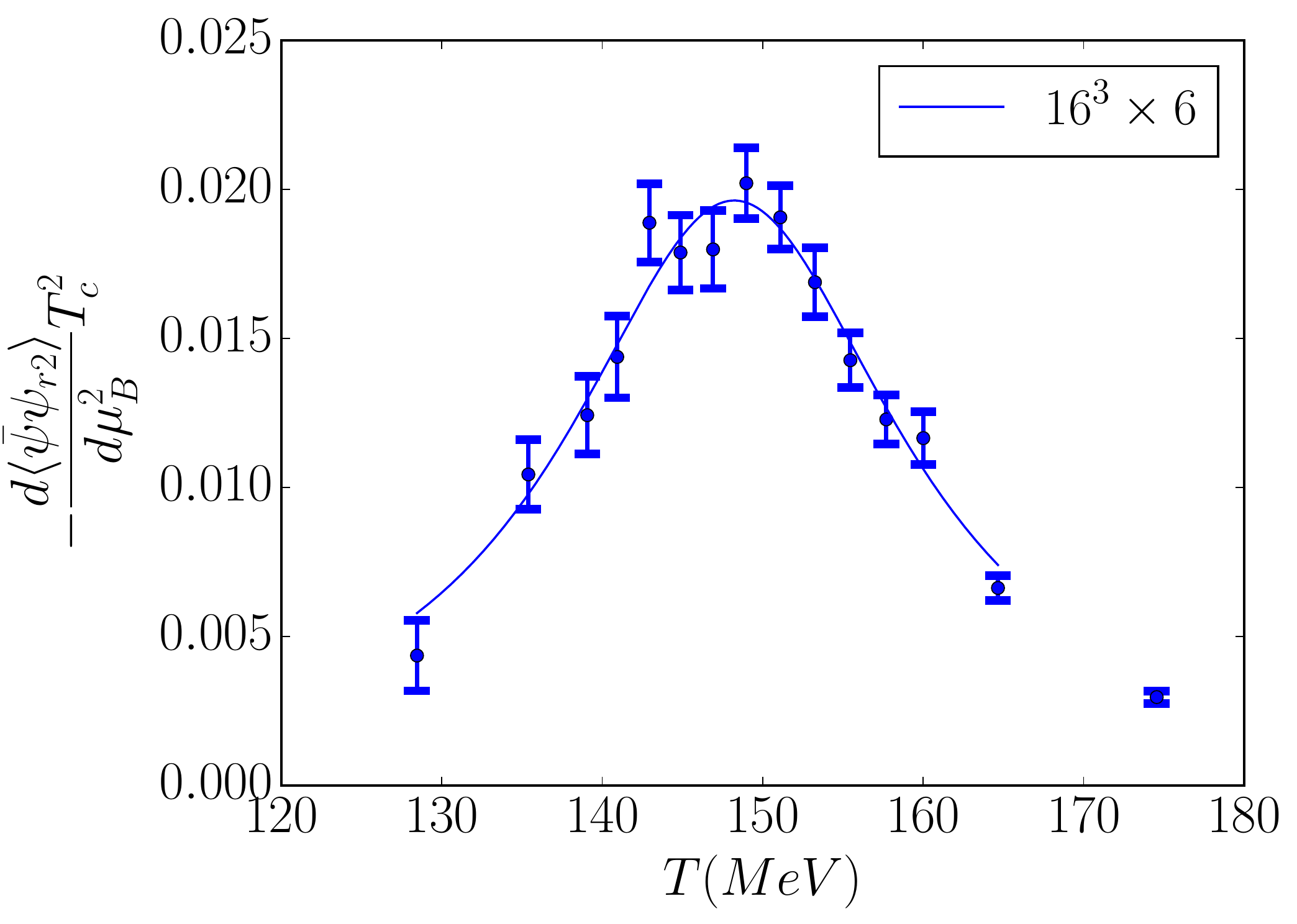}
\includegraphics[scale=0.38]{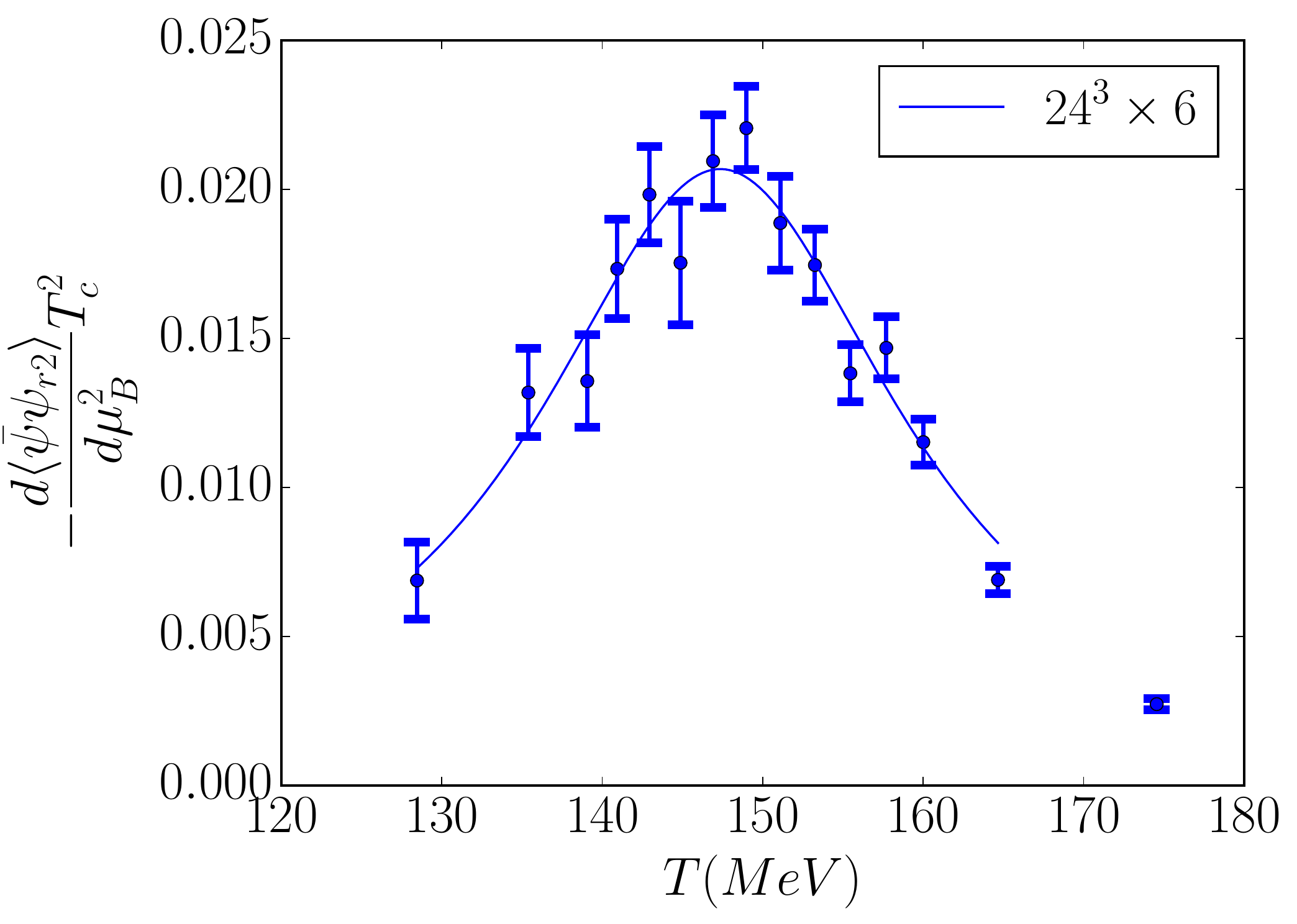}
\includegraphics[scale=0.38]{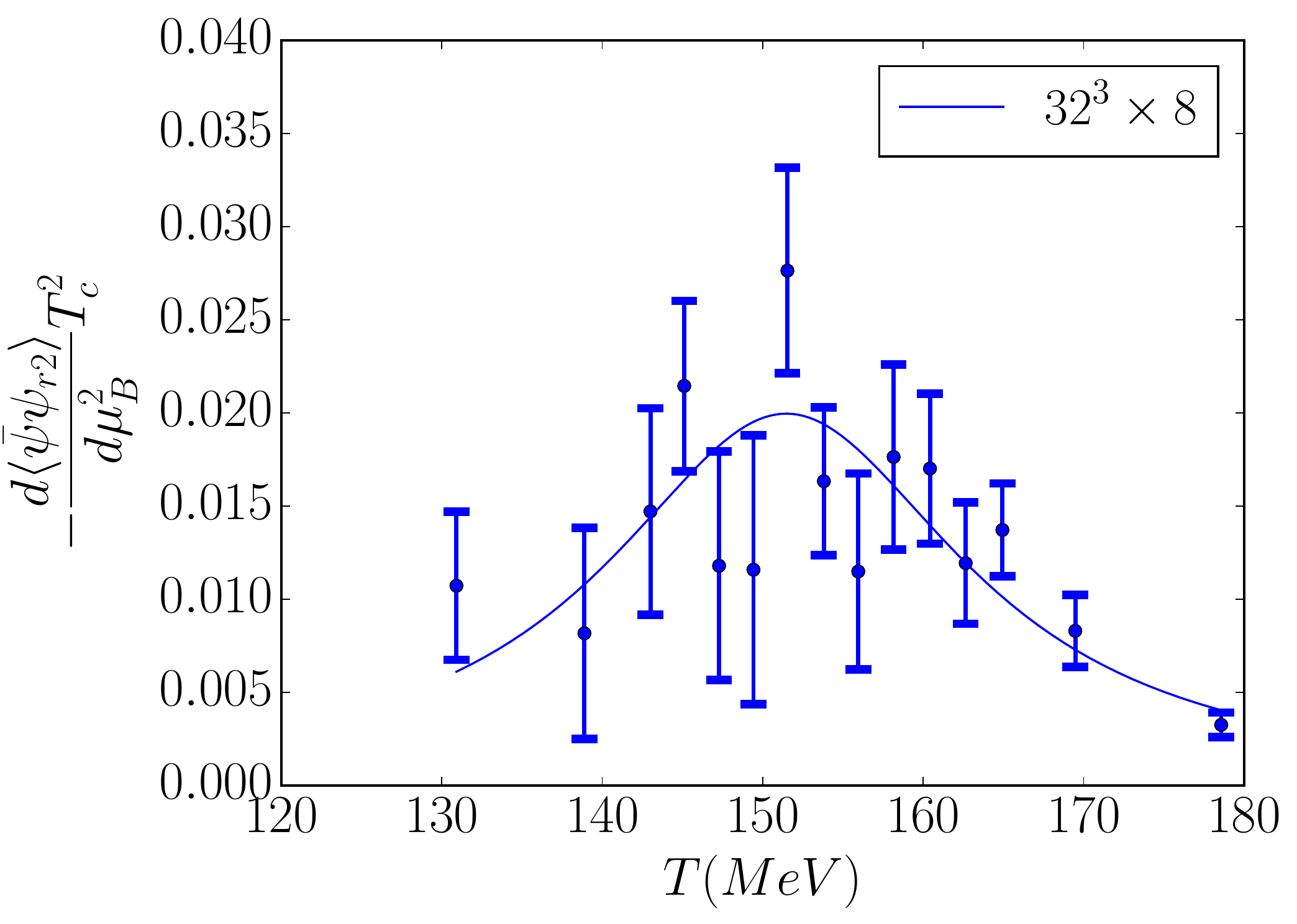}
\caption{First derivative with respect to $\mu_B^2$ of $\langle \bar{\psi}\psi_{r2} \rangle$ for $N_t = 6$ and $N_t = 8$ lattices.}
\label{fig:ffii_renorm2_all}
\end{figure}

\section{Numerical Results}
\label{results}

We performed numerical simulations on four different lattices, with dimensions $16^3 \times 6$, $24^3 \times 6$, $32^3 \times 8$
and $40^3 \times 10$, and adopting the bare parameters reported 
in Table~\ref{tab:lcp} in order to stay on a line of constant physics.
The scale determination is affected 
by an overall systematic error of the order of 
2-3~\%~\cite{tcwup1,befjkkrs}, which however is not relevant to
our final results, which are based on the determination of dimensionless
ratios of quantities measured 
at the critical temperature.
We have adopted the standard
Rational Hybrid
Monte-Carlo algorithm~\cite{rhmc1, rhmc2, rhmc3} 
implemented in two different codes, one running on standard
clusters (NISSA), 
the other on GPUs (OpenStaPLE~\cite{ferrarapisa1,ferrarapisa2})
and developed in OpenACC starting from previous
GPU implementations~\cite{incardona}.

A total of $\approx 15$-$20$ $K$ and $\approx 6$ $K$ Molecular Dynamics 
trajectories have been generated, for each value of $\beta$, 
respectively on the two coarsest lattices
and on the $N_t = 8$ lattice. On the $N_t = 10$ lattice, the derivative of the chiral condensate has been measured exploiting a dedicated high-statistics
simulation performed at the critical temperature and consisting of 
$\approx 100$ $K$ trajectories, while 
for the values of the  chiral condensate around $T_c$ we 
have relied on results obtained in Ref.~\cite{corvo2}.
Also for the values of the chiral condensate 
at zero temperature, which are needed to obtain renormalized quantities, 
we have used results obtained in previous studies~\cite{crow,corvo2}.
Finally, all traces needed in our computations (see Eq.~(\ref{deftraces}))
have been estimated, respectively for the $N_t=6,8,10$ lattices,
every 10, 20 and 50 trajectories by means of noisy estimators,
adopting $256$, $512$ and $512$ random vectors for each flavor and 
for each configuration.

Numerical results obtained on $N_t = 6,8$ for the chiral 
condensate and for its derivative with respect to $\mu_B^2$ are
reported, for both renormalization procedures, respectively
in Figs.~\ref{fig:ff_renorm1_all}, {\ref{fig:ff_renorm2_all}
and in Figs.~\ref{fig:ffii_renorm1_all}, \ref{fig:ffii_renorm2_all}.
As for the derivative, we report in all cases 
the dimensionless combination
$T_c^2 B = T_c^2\, {\partial \langle \bar{\psi}\psi \rangle_r} / {\partial (\mu_B^2)}
$. Statistical errors have been estimated by a jackknife analysis.

\begin{table}[htbp]
\begin{center}
    \begin{tabular}{|c|c|}
    \hline
    Symbol & Fit function \\
    \hline
     A & Atan \\
     T & Tanh \\
     C & Cubic \\
    \hline
     L & Lorentzian \\
     P & Parabola \\
     S & Spline \\
     \hline
    \end{tabular}
\end{center}
\caption {Summary of the 
symbols used in the following tables to describe the ansatz 
for the fitting function.}
\label{tab:symbols_fitfunc}
\end{table}

\begin{table}[t!]
\begin{center}
    \begin{tabular}{c}
{  $16^3 \times 6$ } \\
    \end{tabular}  
    \begin{tabular}{|c|c|c|c|c|}
    \hline
    Fit & $A'(T_c)_{r1}$ & $A'''(T_c)_{r1}$ & $A'(T_c)_{r2}$ & $A'''(T_c)_{r2}$ \\
    \hline
    A & $-0.0258(4)$ & $2.65(44) \cdot 10^{-4}$ & $-0.01088(18)$ & $1.11(18) \cdot 10^{-4}$ \\
    T & $-0.0255(4)$ & $2.14(38) \cdot 10^{-4}$ & $-0.01074(18)$ & $0.90(16) \cdot 10^{-4}$ \\
    C & $-0.0249(5)$ & $1.45(30) \cdot 10^{-4}$ & $-0.01050(20)$ & $0.61(12) \cdot 10^{-4}$ \\
    \hline
    \end{tabular}
    \begin{tabular}{c}
{   $24^3 \times 6$ } \\
    \end{tabular}  
    \begin{tabular}{|c|c|c|c|c|}
    \hline
    Fit & $A'(T_c)_{r1}$ & $A'''(T_c)_{r1}$ & $A'(T_c)_{r2}$ & $A'''(T_c)_{r2}$ \\
    \hline
    A & $-0.0269(3)$ & $2.85(27) \cdot 10^{-4}$ & $-0.01136(11)$ & $1.20(11) \cdot 10^{-4}$ \\
    T & $-0.0265(3)$ & $2.27(25) \cdot 10^{-4}$ & $-0.01116(11)$ & $0.96(10) \cdot 10^{-4}$ \\
    C & $-0.0265(3)$ & $1.86(27) \cdot 10^{-4}$ & $-0.01115(15)$ & $0.78(10) \cdot 10^{-4}$ \\
    \hline
    \end{tabular}  
    \begin{tabular}{c}
{  $32^3 \times 8$ } \\
    \end{tabular}  
    \begin{tabular}{|c|c|c|c|c|}
    \hline
    Fit & $A'(T_c)_{r1}$ & $A'''(T_c)_{r1}$ & $A'(T_c)_{r2}$ & $A'''(T_c)_{r2}$ \\
    \hline
    A & $-0.0226(5)$ & $1.72(30) \cdot 10^{-4}$ & $-0.00993(21)$ & $0.87(15) \cdot 10^{-4}$ \\
    T & $-0.0225(4)$ & $1.50(26) \cdot 10^{-4}$ & $-0.00983(21)$ & $0.74(14) \cdot 10^{-4}$ \\
    C & $-0.0219(6)$ & $1.07(26) \cdot 10^{-4}$ & $-0.00957(28)$ & $0.51(14) \cdot 10^{-4}$ \\
    \hline
    \end{tabular}
\end{center}
\caption {Values obtained for $A'(T_c)$ and $A'''(T_c)$  on 
lattices with $N_t = 6$ and 8. 
Indexes $r1$ and $r2$ refer to the two different definitions
of the renormalized condensate. Derivatives have been take with respect to
the physical temperature, therefore they are reported
in MeV$^{-1}$ and MeV$^{-3}$ units respectively 
for $A'$ and $A'''$.}
\label{tab:A_all}
\end{table}

\begin{table}[t!]
\begin{center}
    \begin{tabular}{c}
{  $16^3 \times 6$ } \\
    \end{tabular}  
    \begin{tabular}{|c|c|c|c|c|}
    \hline
    Fit & $T_c^2\, B(T_c)_{r1}$ & $T_c^2\, 
B''(T_c)_{r1}$ & $T_c^2\, B(T_c)_{r2}$ & 
$T_c^2\, B''(T_c)_{r2}$ \\
    \hline
    L & $0.0467(12)$ & $-6.1(7) \cdot 10^{-4}$ & $0.0197(5)$ & $-2.55(29) \cdot 10^{-4}$ \\
    P & $0.0454(13)$ & $-3.7(6) \cdot 10^{-4}$ & $0.0192(5)$ & $-1.57(27) \cdot 10^{-4}$ \\
    S & $0.0460(12)$ & $-4.9(6) \cdot 10^{-4}$ & $0.0194(5)$ & $-2.05(25) \cdot 10^{-4}$ \\
    \hline
    \end{tabular}
    \begin{tabular}{c}
{  $24^3 \times 6$ } \\
    \end{tabular}  
    \begin{tabular}{|c|c|c|c|c|}
    \hline
    Fit & $T_c^2\, B(T_c)_{r1}$ & $T_c^2\, 
B''(T_c)_{r1}$ & $T_c^2\, B(T_c)_{r2}$ & 
$T_c^2\, B''(T_c)_{r2}$ \\
    \hline
    L & $0.0487(16)$ & $-5.0(8) \cdot 10^{-4}$ & $0.0206(7)$ & $-2.1(3) \cdot 10^{-4}$ \\
    P & $0.0479(17)$ & $-3.9(8) \cdot 10^{-4}$ & $0.0202(7)$ & $-1.7(4) \cdot 10^{-4}$ \\
    S & $0.0485(17)$ & $-4.9(8) \cdot 10^{-4}$ & $0.0205(7)$ & $-2.0(3) \cdot 10^{-4}$ \\
    \hline
    \end{tabular}
    \begin{tabular}{c}
{  $32^3 \times 8$ } \\
    \end{tabular}  
    \begin{tabular}{|c|c|c|c|c|}
    \hline
    Fit & $T_c^2\, B(T_c)_{r1}$ & $T_c^2\, 
B''(T_c)_{r1}$ & $T_c^2\, B(T_c)_{r2}$ & 
$T_c^2\, B''(T_c)_{r2}$ \\
    \hline
    L & $0.044(5)$ & $-3.2(1.8) \cdot 10^{-4}$ & $0.0187(21)$ & $-1.3(7) \cdot 10^{-4}$ \\
    P & $0.041(4)$ & $-1.6(0.9) \cdot 10^{-4}$ & $0.0172(18)$ & $-0.7(4) \cdot 10^{-4}$ \\
    S & $0.042(6)$ & $-2.9(2.1) \cdot 10^{-4}$ & $0.0175(25)$ & $-1.2(9) \cdot 10^{-4}$ \\
    \hline
    \end{tabular}
\end{center}
\caption {Values obtained for $T_c^2\, B(T_c)$ and $T_c^2\, 
B''(T_c)$ on
lattices with $N_t = 6$ and 8. 
Indexes $r1$ and $r2$ refer to the two different definitions
of the renormalized condensate. 
$T_c^2\, B(T_c)$ is dimensionless while 
$T_c^2\, B''(T_c)$ is given in 
MeV$^{-2}$ units.}
\label{tab:B_all}
\end{table}

Reported results have been fitted in order to obtain 
the quantities $A'$, $A'''$, $B$ and $B''$ computed at the 
pseudo-critical temperature needed to determine 
$\kappa$, where derivatives are taken with respect to the 
temperature (see Eqs.~(\ref{defkappa2}) and (\ref{defkappa1})):
\beq
\kappa_1 = \frac{1}{T_c} \frac{T_c^2 B}{A'} \ \ ; \ \ \ \ 
\kappa_2 = \frac{1}{T_c} \frac{T_c^2 B''}{A'''} \, .
\label{summarydefskappa}
\eeq
In order to estimate the systematic uncertainty related to the choice 
of the fitting function, we have tried different ansatzes which are summarized
in Table~\ref{tab:symbols_fitfunc}. 
In particular, for the renormalized condensate we have 
adopted an arctangent, $A = P_1 + P_2\mbox{ }atan (P_3 (T - T_c))$,
an hyperbolic tangent and a cubic polynomial, while for
its $\mu_B^2$ derivative we have considered
a Lorentzian function
$T_c^2 B = \frac{P_1}{P_2^2 + (T-P_3)^2}$, a parabola and a cubic spline.
All best fit results for $A'$, $A'''$, $T_c^2 B$ and $T_c^2 B''$ evaluated 
at $T_c$ are reported 
in Tables~\ref{tab:A_all} and \ref{tab:B_all}: reported
errors include the systematic one related to the choice of the 
fit range. Results obtained for the pseudocritical temperature 
$T_c$ are instead reported in Table~\ref{tab:Tc_all_lattices}.

In principle, the uncertainty in the determination of $T_c$
should contribute to the error given for $A'$, $A'''$, $B$ and $B''$
computed at $T_c$, however this contribution turns out to be negligible
in most cases,
apart from the $40^3 \times 10$ (and in particular for $B$)
where it is marginally 
appreciable because of the larger uncertainty on $T_c$.
For the determinations of $T_c$ and $A'$ 
on the $N_t=10$ lattice
we reused the data already obtained in Ref.~\cite{corvo2},
while $B$ has been obtained from the single dedicated simulation
performed at $T \simeq T_c$. Results for $A'$ and $B$ are 
reported respectively in Tables~\ref{tab:A_40x40x40x10} and 
\ref{tab:B_40x40x40x10}; the estimate for the error on 
$B (T_c)$ stemming from the uncertainty on $T_c$
has been based on the data available for $B$ as a function of 
$T$ on the $24^3 \times 6$ lattice.

\begin{table}[h!]
\begin{center}
    \begin{tabular}{|c|l|l|}
    \hline
    Lattice & $T_c(\bar{\psi}\psi_{r1})$ & $T_c(\bar{\psi}\psi_{r2})$ \\
    \hline
    $16^3 \times 6$ & $148.2(4)$ & $148.3(4)$ \\
    $24^3 \times 6$ & $149.1(2)$ & $149.2(2)$ \\
    $32^3 \times 8$ & $154.4(4)$ & $154.7(4)$ \\
    $40^3 \times 10$ & $154.7(1.6)$ & $154.4(1.6)$ \\
    \hline
    \end{tabular}
\end{center}
\caption {Values obtained for the critical temperature from the fits.
Reported errors take into account the systematic uncertainty 
related to the choice of the fitting function and range, but not  
the overall uncertainty on the determination of the physical
scale, which is of the order of $2-3$ \%~\cite{tcwup1,befjkkrs}.
Values reported for the $40^3 \times 10$ lattice
are based on results reported in Ref.~\cite{corvo2}.}
\label{tab:Tc_all_lattices}
\end{table}

\begin{table}[t!]
\begin{center}
    \begin{tabular}{|c|c|c|c|c|}
    \hline
    Fit & $A'(T_c)_{r1}$ & $A'(T_c)_{r2}$\\
    \hline
    A & $-0.0231(09)$ & $-0.0093(4)$\\
    T & $-0.0226(11)$ & $-0.0091(4)$\\
    C & $-0.0215(12)$ & $-0.0089(5)$\\
    \hline
    \end{tabular}
\end{center}
\caption {Values obtained for $A'(T_c)$ on the $40^3 \times 10$ lattice,
based on results obtained in Ref.~\cite{corvo2}. Units and conventions are
as in Table~\ref{tab:A_all}.
}
\label{tab:A_40x40x40x10}
\end{table}

\begin{table}[h!]
\begin{center}
    \begin{tabular}{|c|c|}
    \hline  
    $T_c^2\, B(T_c)_{r1}$ & $T_c^2\, B(T_c)_{r2}$\\
    \hline
    $0.052(6)(2)$ & $0.0217(25)(10)$\\
    \hline
    \end{tabular}
\end{center}
\caption {Values obtained for $T_c^2 B(T_c)$ on the $40^3 \times 10$ lattice.
The second error refers to the uncertainty
in the determination of the pseudo-critical temperature at $\mu_B = 0$.}
\label{tab:B_40x40x40x10}
\end{table}

As it can be appreciated from Tables~\ref{tab:A_all} and 
\ref{tab:B_all}, the systematic uncertainties related to the 
choice of the fitting function are in a few  cases
comparable or larger than statistical errors.
For this reason, in order to obtain our final estimates
for $\kappa_1$ and $\kappa_2$, which are 
are based on Eq.~(\ref{summarydefskappa}) and are 
reported in 
Table~\ref{tab:k_Nt_all},
we have considered the dispersion of values corresponding
to all possible combinations of different fitting functions,
and added it, when appreciable, to the statistical error.
As one can see, 
present statistics are not enough to reach reliable estimates 
of $B''$ on lattices with $N_t > 6$, where they are affected 
by errors of the order of 50~\%.

\begin{table}[htbp]
\begin{center}
    \begin{tabular}{|c|c|c|c|c|}
    \hline
    Lattice & $\kappa_1(\bar{\psi}\psi_{r1})$ & $\kappa_2(\bar{\psi}\psi_{r1})$ & $\kappa_1(\bar{\psi}\psi_{r2})$ & $\kappa_2(\bar{\psi}\psi_{r2})$ \\
    \hline
    $16^3 \times 6$ & $0.0122(5)$ & $0.016(6)$ & $0.0122(5)$ & $0.016(6)$\\
    $24^3 \times 6$ & $0.0122(4)$ & $0.015(4)$ & $0.0122(4)$ & $0.015(4)$\\
    $32^3 \times 8$ & $0.0126(14)$ & $0.014(9)$ & $0.0121(13)$ & $0.012(8)$ \\
    $40^3 \times 10$ & $0.0146(19)$ & - & $0.0154(21)$ & - \\
    \hline
    \end{tabular}
\end{center}
\caption {Curvature coefficient $\kappa$ obtained for different definitions
and lattice sizes.}
\label{tab:k_Nt_all}
\end{table}

\subsection{Discussion of results on 
$N_t = 6$ lattices: finite size effects and   
comparison between $\kappa_1$ and $\kappa_2$}
\label{res_Nt_6}

Results obtained on $N_t = 6$ for different spatial sizes permit
us to make an assessment of the relevance of finite size effects.
For a closer comparison, in Fig.~\ref{fig:ffii_renorm12_16_vs_24}
we plot together results obtained for $T_c^2 B$ on the different
volumes.
A mild dependence on the spatial volume is visible for some quantities,
like for instance the pseudo-critical temperature. However results 
obtained for $\kappa$ on the $16^3 \times 6$ and
$24^3 \times 6$ lattices are in perfect agreement within errors,
thus confirming the small sensitivity of $\kappa$ to finite
size effects already observed in studies exploiting 
analytic continuation~\cite{crow}.

Regarding the comparison between $\kappa_1$ and $\kappa_2$,
one observes a slight tendency for $\kappa_2$ to be larger than
$\kappa_1$, however this is not significant within errors, 
which are of the order of 30\% for $\kappa_2$; therefore 
the two determinations are compatible within our present level 
of accuracy.
Finally, no significant difference is observed between determinations
obtained with the two different renormalization prescriptions for 
the chiral condensate.

\begin{figure}[t!]
\centering
\includegraphics[scale=0.38]{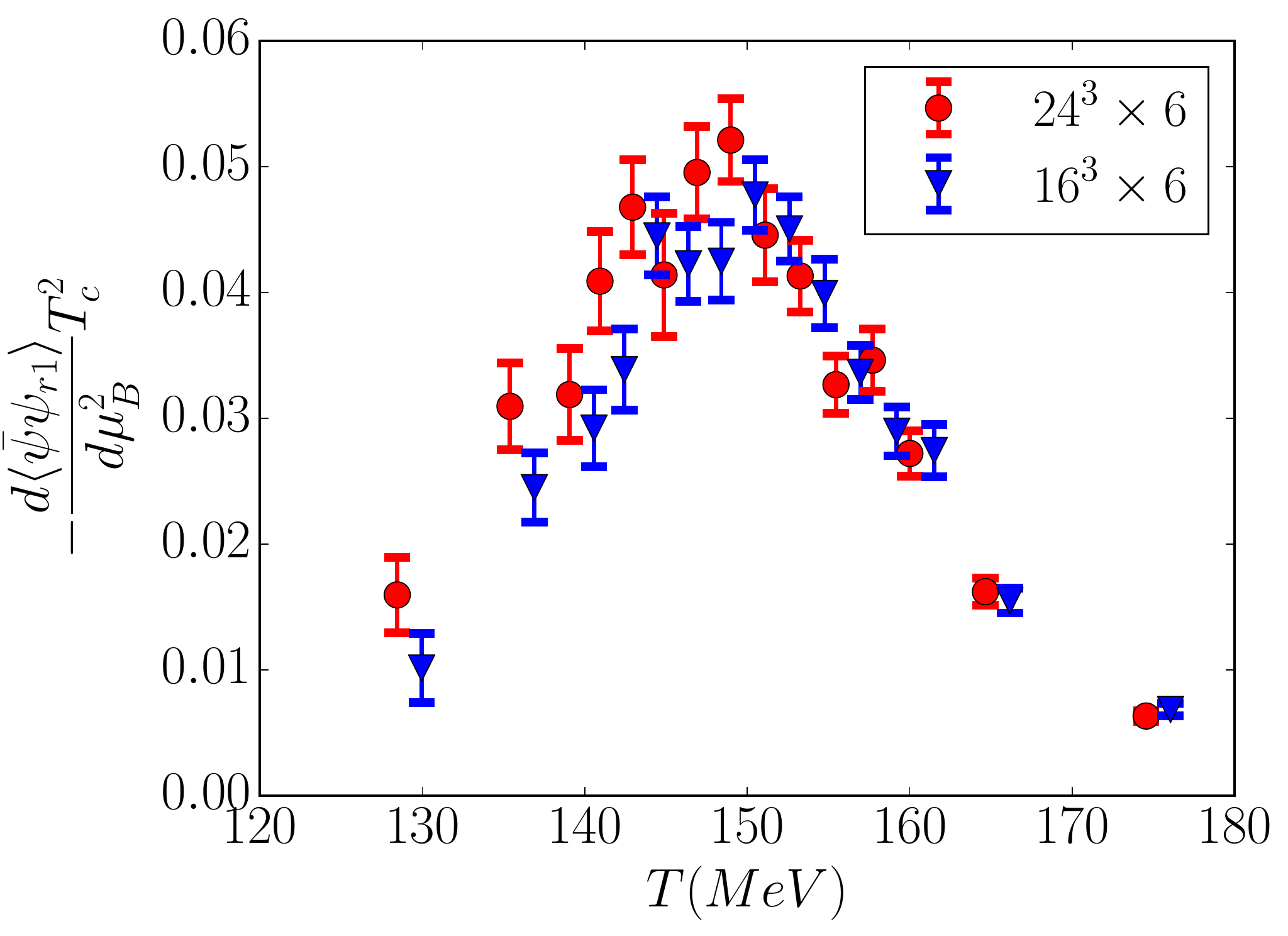}
\includegraphics[scale=0.38]{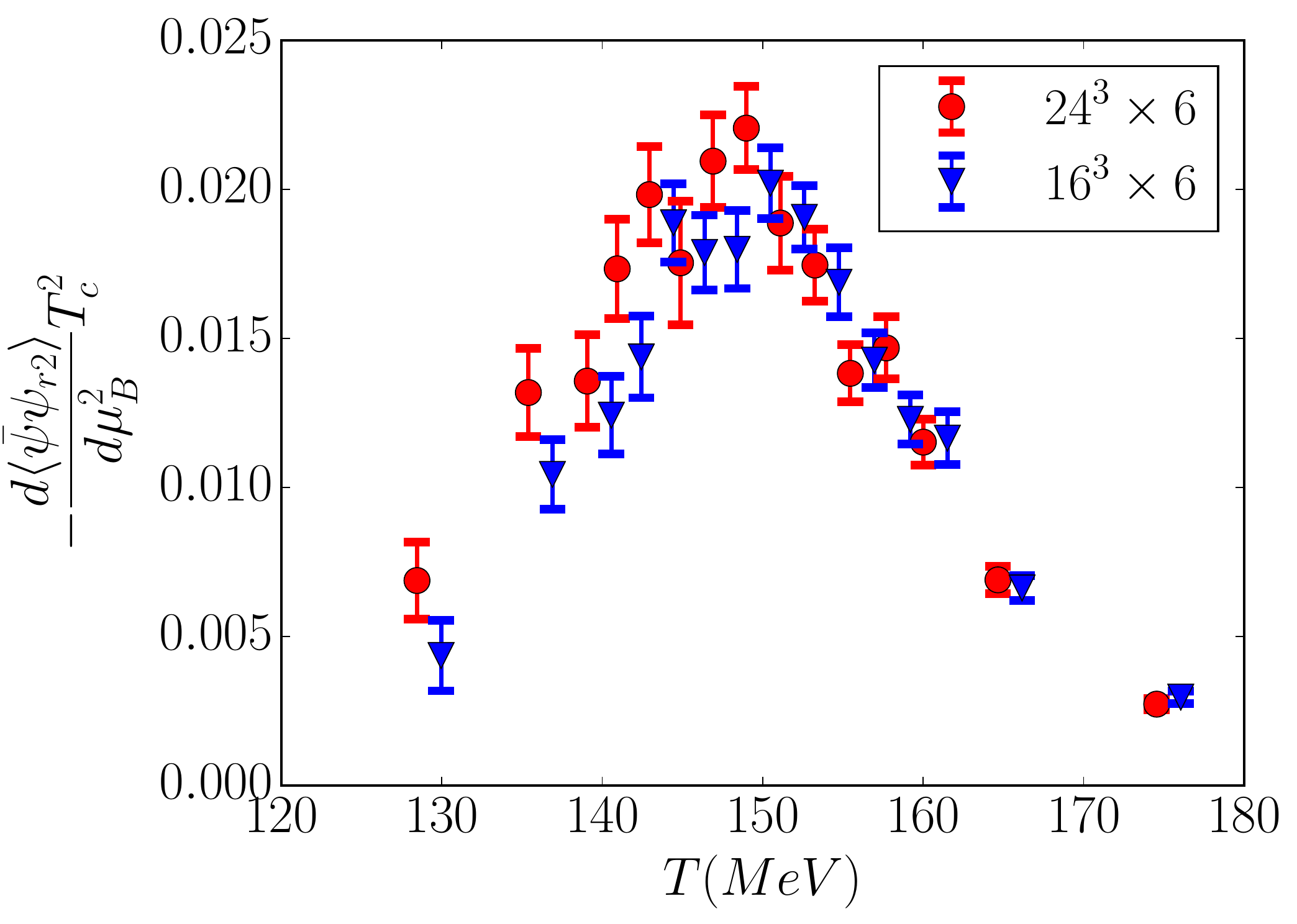}
\caption{First derivative with respect to $\mu_B^2$ of the renormalized condensate: comparison between the results obtained on the $16^3$ and the $24^3$ lattices for the two different definitions. Data points have been slightly 
shifted horizontally to improve readability.}
\label{fig:ffii_renorm12_16_vs_24}
\end{figure}

\subsection{Continuum extrapolation of $\kappa_1$ and comparison with analytic continuation} 
\label{res_continuum}

A continuum extrapolation is presently possible only for
$\kappa_1$, for which results are available for three
different lattice spacings, corresponding to $N_t =$ 6, 8 and 10,
while for $\kappa_2$ we can just say that no appreciable variations 
are observable going from $N_t = 6$ to 
$N_t = 8$, however errors for $N_t = 8$ are too large to make this 
statement of any significance.
Assuming corrections linear in $a^2$, $\kappa_1(a^2) = 
\kappa_1^{cont} + O(a^2)$, and since $T = \frac{1}{N_t a}$,
an extrapolation to the continuum limit for $\kappa_1$ 
can be obtained by a best fit of the function
\begin{equation}
    \kappa(N_t) = \kappa^{cont} + A \frac{1}{N_t^2} \mbox{ . }
\end{equation}
The values of $\kappa_1$ obtained on the 
$24^3 \times 6$, $32^3 \times 8$ and $40^3 \times 10$ lattices are
reported in Fig.~\ref{fig:k_continuum}, where are also illustrated
the results of the continuum extrapolation,
which gives back $\kappa_1^{cont}(\bar{\psi}\psi_{r1}) = 0.0147(22)$
and $\kappa_1^{cont} (\bar{\psi}\psi_{r2}) = 0.0144(26)$,
with a reduced chi-squared respectively $0.42$ and $1.38$.

As a final estimate, we quote the average value
$\kappa_1 = 0.0145(25)$, which is in very good
agreement within errors with previous results obtained 
via analytic continuation, in particular $\kappa = 0.0135(20)$
from Ref.~\cite{corvo2} where
the same discretization and numerical setup
of chemical potentials have been adopted.

\begin{figure}[h!]
\centering
\includegraphics[scale=0.4]{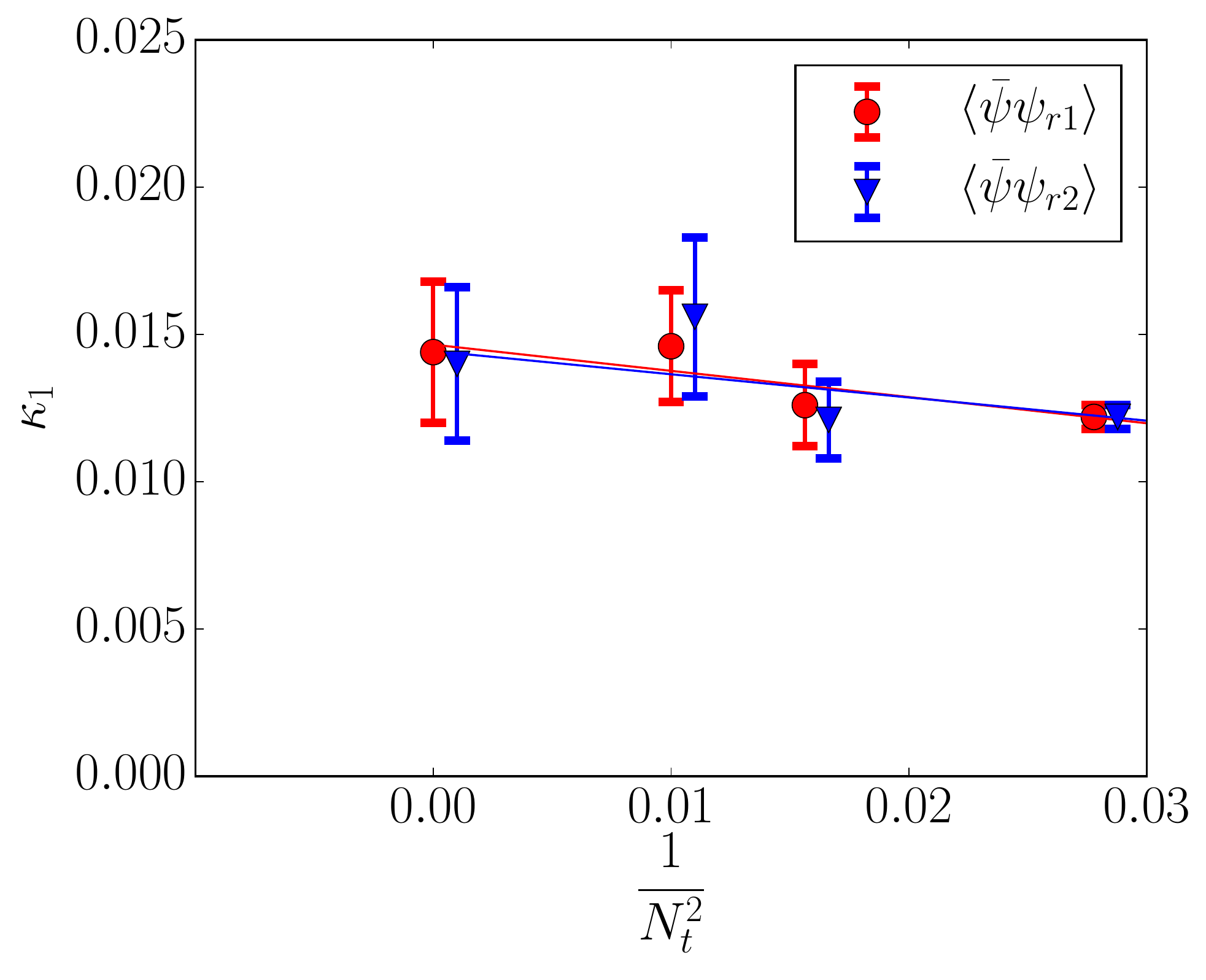}
\caption{Values obtained for $\kappa_1$ 
on different values of $N_t$ and continuum extrapolation.}
\label{fig:k_continuum}
\end{figure}

Apart from the continuum limit, a direct comparison with analytic continuation
can be made separately for the different lattice sizes and for 
the different definitions of $\kappa$. This is possible in particular
exploiting the results reported in Ref.~\cite{crow}, where
values of both $\kappa_1$ and $\kappa_2$ 
have been reported for  
$16^3 \times 6$, $24^3 \times 6$ and $32^3 \times 8$ lattices
and for both renormalizations of the chiral condensate.
An inspection of Table~\ref{tab:k_comparison}, where results
from this work and from Ref.~\cite{crow} are reported together
for lattices where both are available, reveals some tension 
between Taylor expansion and analytic continuation
on the coarsest lattices, which tends to disappear 
on the finer lattices, also because of the larger statistical 
errors.

\begin{table}[h!]
\begin{center}
    \begin{tabular}{|c|c|c|c|c|c|}
    \hline
    $Work$ & $Lattice$ & $\kappa_1(\bar{\psi}\psi_{r1})$ & $\kappa_2(\bar{\psi}\psi_{r1})$ & $\kappa_1(\bar{\psi}\psi_{r2})$ & 
$\kappa_2(\bar{\psi}\psi_{r2})$ \\
    \hline
    This             & $16^3 \times 6$ & $0.0122(5)$ & $0.016(6)$ & $0.0122(5)$ & $0.016(6)$\\
    Work                   & $24^3 \times 6$ & $0.0122(4)$ & $0.015(4)$ & $0.0122(4)$ & $0.015(4)$\\
                       & $32^3 \times 8$ & $0.0126(14)$ & $0.014(9)$ & $0.0121(13)$ & $0.012(8)$\\
    \hline
    \cite{crow}        & $16^3 \times 6$ & $0.0136(3)$ & $0.0133(4)$ & $0.0124(3)$ & $0.0133(5)$\\
                       & $24^3 \times 6$ & $0.0139(3)$ & $0.0150(7)$ & $0.0147(3)$ & $0.0152(7)$\\
                       & $32^3 \times 8$ & $0.0136(3)$ & $0.0142(7)$ & $0.0131(3)$ & $0.0135(7)$\\
    \hline
    \end{tabular}
\end{center}
\caption {Comparison of results obtained for different 
definitions of $\kappa$ in this work with those reported in Ref.~\cite{crow}
via analytic continuation.}
\label{tab:k_comparison}
\end{table}

\section{Discussion and Conclusions}
\label{conclusions}

Recently, many numerical investigations have been carried out to determine the curvature $\kappa$ of the pseudocritical line in the QCD phase diagram
departing from the $\mu_B = 0$ axis. 
Estimates obtained by the Taylor expansion technique have been
generally lower than those obtained by analytic continuation, however, 
since the transition is a crossover, care is needed when comparing results 
obtained by studying different observables.

In this work, the curvature of the pseudocritical line has been studied 
for $N_f = 2 + 1$ QCD,
via Taylor expansion and through numerical simulations performed
using the tree-level Symanzik gauge action and the stout-smeared 
staggered fermion action. This is the same discretization adopted
in Refs.~\cite{Endrodi2011,crow,corvo2};
moreover, we have adopted the same observables
and definitions of $\kappa$ investigated in those previous
studies, in order to make the comparison closer.

In particular, the location of the phase transition has been determined from the inflection point of the chiral
condensate and 
using two renormalization prescriptions, $\bar{\psi}\psi_{r1}$ and $\bar{\psi}\psi_{r2}$, defined respectively in Eqs.~(\ref{rencond})  
and (\ref{eq:ren_pres_wupp}).
The curvature coefficient has been calculated using two
different definitions: the first one, $\kappa_1$, adopted 
in Ref.~\cite{Endrodi2011}, assumes that the value of the renormalized 
condensate stays constant at the critical temperature
as the baryon chemical potential is switched on.
The second one, $\kappa_2$, which is the same adopted 
in Refs.~\cite{crow,corvo2}, looks at how the actual
inflection point of the condensate moves as a function 
of $\mu_B$: it is preferable because it is does not rely on 
particular assumptions, however it involves the computation
of higher order derivatives, leading to larger numerical
uncertainties.

\begin{figure}[t!]
\centering
\includegraphics[scale=0.4]{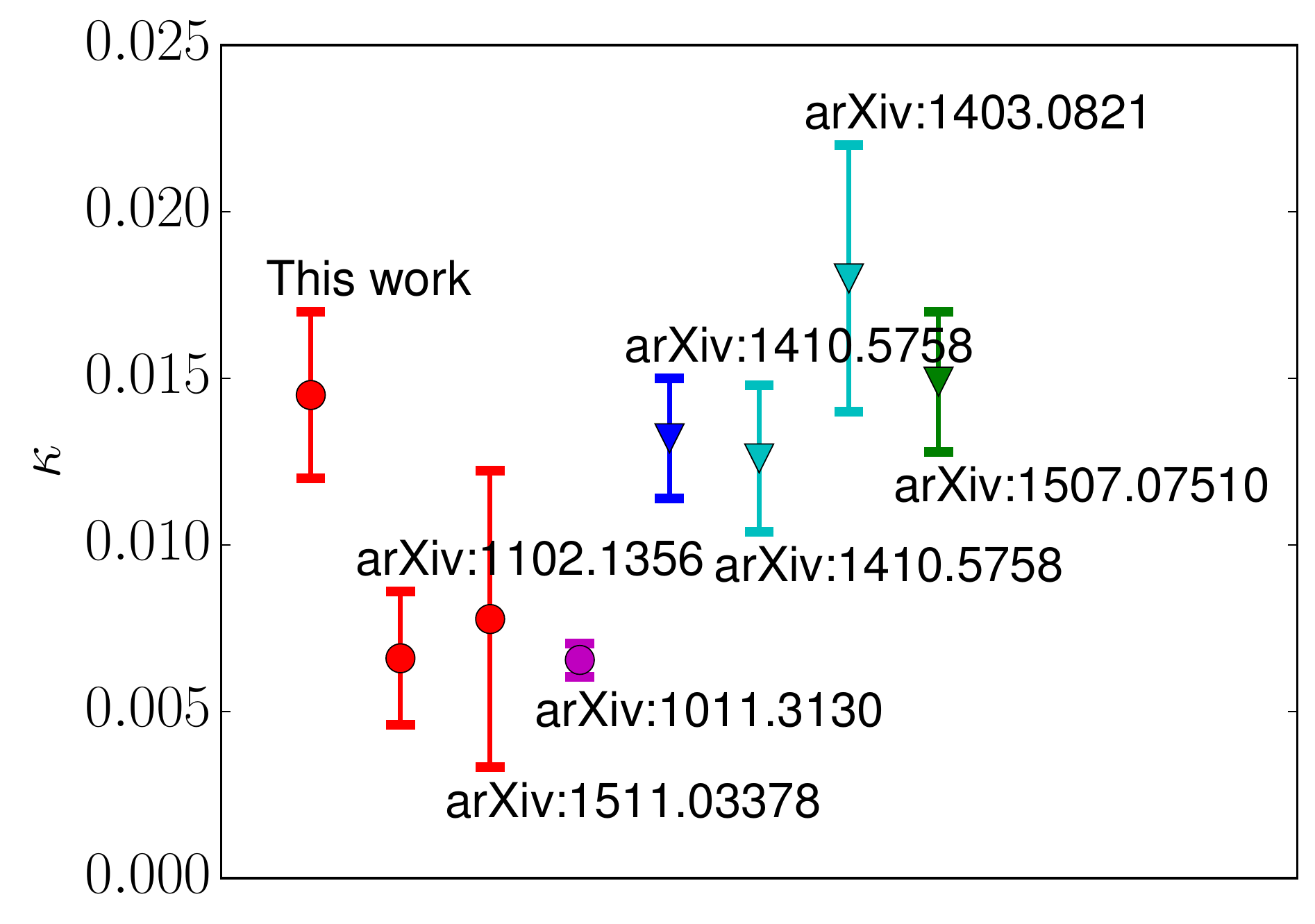}
\caption{Curvature coefficient $\kappa$: comparison with previous determinations.
From the left to the right: estimate by this work, Ref.~\cite{Endrodi2011}, Ref.~\cite{Kaczmarek2011}, Ref.~\cite{crow}, Ref.~\cite{crow}, Ref.~\cite{ccp}
and Ref.~\cite{ntc}. Bars marked with circles and triangles indicate that estimates have been obtained respectively by Taylor expansion
and analytic continuation. More precisely, the red, magenta, blue, cyan and green colors indicate that the estimate has been obtained respectively by Taylor
expansion $+$ chiral condensate, Taylor expansion $+$ chiral susceptibility, analytic continuation $+$ chiral condensate,
analytic continuation $+$ chiral susceptibility and analytic continuation $+$ combined analysis of various observables.}
\label{fig:k_comparison}
\end{figure}

Our main results can be summarized as follows:

\begin{enumerate}

\item No statistically significant effect due to the renormalization prescription has been observed and finite size effects have been
      found to be negligible. 
The values obtained for $\kappa_2$ are generally higher than the values obtained for $\kappa_1$,
      however the difference is well within statistical errors, 
therefore 
the two determinations are compatible with our present level 
of accuracy.

\item The values obtained for $\kappa_2$ 
are compatible with those obtained in Ref.~\cite{crow} via analytic 
continuation, 
however present statistics are not enough 
to determine a reliable estimate for 
$\kappa_2$ via Taylor expansion 
on finer lattices and to perform an extrapolation 
to the continuum.
Overall, no discrepancy is observed between the results obtained 
by Taylor expansion and those obtained by analytic continuation also
for $\kappa_1$, though some tension is present 
on the coarsest lattices, which however 
tends to disappear on finer lattices.

\item
The final continuum extrapolation that we have given for 
$\kappa_1$ is 
$\kappa^{cont} = 0.0145(25)$,
which is in agreement 
with results from analytic continuation~\cite{crow,ccp,ntc}.
In Fig.~\ref{fig:k_comparison} we report a summary 
of the most recent determinations of $\kappa$ obtained
for QCD at or close to the physical point: the possible
tension between analytic continuation and 
earlier results obtained via Taylor expansion 
seems to disappear, leaving place to a convergence 
of the two methods.

\end{enumerate}

Regarding the tension (slightly above 2$\sigma$) between our present
results and the results reported in Ref.~\cite{Endrodi2011}, where
the same discretization, observables and definition of $\kappa_1$ were
adopted, a possible explanation could be in the different 
way adopted to take the continuum limit. In our study, for each 
$N_t$, we have determined $\kappa_1$ at the 
corresponding pseudo-critical temperature found at $\mu_B = 0$,
and then we have extrapolated those values to 
the $N_t \to \infty$ limit. In Ref.~\cite{Endrodi2011}, instead,
the definition of $\kappa_1$ has been first extended to a wide range
of temperatures around $T_c$, still based on monitoring how
points (temperatures) where the renormalized condensate assumes a
fixed value change as a function of $\mu_B$; then a  
continuum extrapolation has been performed over the whole range,
thereafter taking the value of this extrapolated function at 
$T_c$.

Two final considerations are in order. First,
while a proper extrapolation to the continuum limit for $\kappa_2$ is probably out of reach with current computational resources, more statistics would allow to
improve the results and make a closer comparison 
at least on $N_t = 6$ lattices.
Second, it must be stressed that the convergence towards a common continuum
extrapolated value of $\kappa$, which is indicated
by recent determinations from analytic continuation and Taylor 
expansion, is still limited to results obtained only from staggered 
fermion simulations. It would be important, in the future, to have 
confirmations also from studies adopting different fermion discretizations.

\acknowledgments
Numerical simulations have been performed on the MARCONI
and GALILEO machines at CINECA, based on the 
agreement between INFN and CINECA (under project INF17\_npqcd
and INF18\_npqcd), on the COKA cluster at INFN-Ferrara
and at the Scientific Computing
Center at INFN-Pisa. 
FN acknowledges financial support from the INFN HPC\_HTC project.

\end{document}